\documentclass[prd,aps,twocolumn,a4paper,showkeys,nofootinbib]{revtex4-1}

\usepackage{graphicx,psfrag}
\usepackage{mathrsfs}
\usepackage{amsmath,amsfonts,amssymb}
\usepackage{multirow}
\usepackage{comment}
\usepackage{hyperref}
\usepackage{enumitem}
\usepackage{tcolorbox}
\usepackage[utf8]{inputenc} 
\usepackage[T1]{fontenc}    
\newcommand{\be}{\begin{equation}}
\newcommand{\ee}{\end{equation}}
\newcommand{\bea}{\begin{eqnarray}}
\newcommand{\eea}{\end{eqnarray}}
\newcommand{\bel}{\begin{align}}
\newcommand{\eel}{\end{align}}

\def\cf{\textit{Cf.}}
\def\ie{\textit{i.e.}}

\def\p{\partial}

\def\Le{L_{\rm e}}

\def\gccm{{\rm g\,cm^{-3}}}
\def\Msun{{\rm M_{\odot}}}
\def\Mo{\Msun}

\def\GMc2{{\rm G M_{\odot} c^{-2}}}

\usepackage{pifont} 
\usepackage{color}
\definecolor{cyan}{rgb}{0,0.9,0.9}
\definecolor{orange}{rgb}{0.9,0.5,0}
\definecolor{magenta}{rgb}{1,0,1}
\definecolor{purple}{rgb}{0.8,0.4,0.8}
\definecolor{gray}{rgb}{0.8242,0.8242,0.8242}
\definecolor{light-gray}{gray}{0.95}

\definecolor{mygreen}{RGB}{0,155,0}

\begin{document}

\title{Dynamical stability of Magnetic Supramassive Neutron Stars}

\author{Jorge \surname{L. Rodr{\'i}guez-Monteverde}$^{2,1}$}\email[]{jorge.lopez.rodriguez-monteverde@rai.usc.es}
\author{Sebastiano \surname{Bernuzzi}$^{1}$}\email[]{sebastiano.bernuzzi@uni-jena.de}
\author{William \surname{Cook}$^{1,3}$}
\author{Maximilian \surname{Jacobi}$^{1}$}
\author{Brynmor \surname{Haskell}$^{4,5}$}

\affiliation{${}^1$Theoretisch-Physikalisches Institut, Friedrich-Schiller-Universit{\"a}t Jena, 07743, Jena, Germany}
\affiliation{${}^2$Instituto Galego de Física de Altas Enerxías, Universidade de
Santiago de Compostela, 15782 Santiago de Compostela, Galicia, Spain}
\affiliation{${}^3$ Mathematical Sciences and STAG Research Centre,
University of Southampton, Southampton SO17 1BJ, United Kingdom}
\affiliation{${}^4$INFN, Sezione di Milano, Via Celoria 16, 20133, Milano, Italy}
\affiliation{${}^5$Dipartimento di Fisica, Università di Milano, Via Celoria 16, 20133, Italy}

\date{\today}

\begin{abstract}
We investigate the dynamical stability of strongly
magnetized supramassive neutron stars close to the mass-shedding limit
using fully general-relativistic magnetohydrodynamic simulations in
three spatial dimensions. Equilibrium configurations are constructed
using a microphysical equation of state and considering poloidal
magnetic fields up to maximum strengths ${\sim} 10^{17}\,\mathrm{G}$. 
The initially unstable magnetic fields reorganize into mixed
poloidal--toroidal configurations on Alfv\'en timescales shorter than 
the stellar rotation period. The subsequent dynamical evolution is
governed by timescales set by the Lehnert number.
For stars with masses sufficiently lower than the maximum mass, 
magnetic stresses redistribute angular momentum and drive the
development of differential rotation, 
leading to a new quasi-stationary configuration that remains
dynamically stable over several rotation periods.
The outer layers of these stars are subject to magnetorotational
instability.
Magnetically-driven winds develop from the outer layers around the
equatorial plane and remain
baryon-matter dominated with magnetization 
${\lesssim}0.2$. Over the simulated timescale, winds carry a total
rest mass of order ${\sim}10^{-3}\mathrm{M}_\odot$ at outflow rates of up to a few
$\Mo/\mathrm{s}$.
For stars close to (or beyond) the maximum-mass turning point along the stellar equilibrium sequence, 
strong magnetic fields trigger gravitational collapse by
redistributing angular momentum away from the stellar core.
The resulting black hole remnants are characterized by 
angular velocities up to ${\sim}49\%$ larger and angular momentum
${\lesssim}8\%$ smaller than those of the initial equilibrium stars.
Our results show that extreme 
magnetic-field instabilities can significantly alter the structure
of supramassive neutron stars, with 
possible implications for magnetars and long-lived
binary neutron-star merger remnants.
\end{abstract}


\maketitle




\section{Introduction}
\label{sec:intro}

Neutron stars are rapidly rotating compact objects that can host
extreme magnetic fields, with intensities reaching
${\sim}10^{12}$--$10^{15}$~G in standard pulsars and magnetars
\cite{Mereghetti:2008je,Kaspi:2017fwg,Wang:2023qww}.
These magnetic fields are thought to originate during the core collapse
of massive stars and/or in the proto-neutron star phase, possibly
through a combination of fossil fields, convective dynamos, and
magnetothermal processes.
Despite decades of observations across the electromagnetic spectrum,
including radio, X-ray, and gamma-ray emission (such as soft gamma-ray
repeaters, anomalous X-ray pulsars, and giant flares
\cite{Hurley:1998ks,Palmer:2005mi,Mereghetti:2023iwx}) the internal
structure and magnetic configuration of neutron stars remain poorly
constrained.

Binary neutron star mergers (BNSM) provide an additional channel for the
formation of strongly magnetized neutron stars. In particular, mergers
with sufficiently low total mass, $M_{\rm bin}\lesssim 1.2-1.35\Mo$, can
produce long-lived remnants endowed with magnetic
fields as large as ${\sim}10^{16}$--$10^{17}$~G, see
e.g.~\cite{Radice:2018xqa,Gutierrez:2026ngt,Kiuchi:2026pgb}.
Angular momentum redistribution driven by magnetic and viscous effects
may drive the remnants' rotation toward uniform rotation on secular timescales
\cite{Radice:2017zta,Shibata:2017jyf}, although the precise mechanisms
remain uncertain. Strong magnetic fields may significantly delay
collapse, potentially sustaining the remnant for timescales of up to
years even in the presence of only moderate rotation
\cite{Suvorov:2021ymy}.

Despite their astrophysical relevance, several aspects of the
structure, stability, and dynamical evolution of strongly magnetized
and rapidly rotating neutron stars remain insufficiently explored.

Supramassive neutron stars (SMNSs) are uniformly rotating equilibrium
configurations with masses exceeding the maximum mass of nonrotating
(Tolman--Oppenheimer--Volkoff, TOV) solutions for a given equation of
state (EOS) \cite{Cook:1992,Cook:1993qr,Beniamini:2021tpy}.
Rotation provides additional support against gravity, allowing masses
up to ${\sim}20\%$ larger than the TOV limit, \ie,
$M \gtrsim 1.2\,M^{\rm TOV}_{\rm max}$.
Studies of equilibrium sequences at fixed baryonic mass show that SMNSs
evolve toward instability as they lose angular momentum, spinning up
toward collapse on secular timescales \cite{Cook:1992}.
Configurations at the mass-shedding limit are dynamically unstable to
quasi-radial perturbations and are expected to undergo gravitational
collapse to a black hole \cite{Shibata:1999yx}.
Finite-temperature effects can temporarily increase the maximum
supported baryonic mass along stable branches, but their impact
diminishes at high central densities near the maximum mass.
Near this limit, SMNSs are less compact and possess more extended
envelopes, reaching the mass-shedding threshold at lower angular
velocities compared to colder configurations \cite{Kaplan:2013wra}.
All of these results have been established in general
relativity in the absence of magnetic fields.

Equilibrium configurations of magnetized neutron stars have been constructed by
solving the coupled Einstein--Maxwell equations, e.g.~
\cite{Bocquet:1995je,Tsokaros:2015fea,Pili:2014npa,Pili:2017yxd}.
These studies indicate that magnetic fields as strong as
${\sim}10^{17}$~G can be supported in equilibrium, consistent with
virial theorem estimates \cite{Chandrasekhar:1953zz,Bocquet:1995je}.
However, the structure and stability of such configurations depend
sensitively on the magnetic field geometry.

It is well established that purely poloidal or purely toroidal magnetic
fields are unstable. Poloidal fields are subject to varicose
instabilities, while toroidal fields suffer from the Tayler 
instability \cite{Tayler:1957a,Tayler:1973a,Markey:1973a,Markey:1974a,
Wright:1973a,Braithwaite:2005su}. Stable configurations are therefore
expected to involve mixed field geometries, often described as
twisted-torus structures \cite{Braithwaite:2005xi,Braithwaite:2005md}.
Time-dependent studies, both in Newtonian gravity and in
general-relativistic magnetohydrodynamics (GRMHD), have explored the
nonlinear development of these instabilities and the possible emergence
of stable equilibria~\cite{Kiuchi:2008ss,Ciolfi:2011xa,Lasky:2011un,Ciolfi:2012en,Sur:2020hwn,Sur:2021awe,Tsokaros:2021pkh,Cheong:2024stz,Cook:2025zzy,Capobianco:2026ots,Li:2026ngu}.
Rapid rotation can partially suppress poloidal-field instabilities on
Alfv\'en timescales \cite{Pinas:2025bpq}, while purely toroidal fields
remain unstable if the magnetic energy becomes sufficiently large
compared to the rotational energy \cite{Kiuchi:2008ss,Kiuchi:2011yt}.
On longer timescales, magnetic field evolution is governed by processes
such as ambipolar diffusion, which can lead to gradual magnetic energy
dissipation \cite{Suvorov:2021ymy}.
Determining the long-term stability of magnetized neutron
stars remains a computationally challenging open problem.

In this work, we investigate the dynamical stability of strongly
magnetized supramassive neutron stars close to the mass-shedding limit using
three-dimensional numerical relativity simulations.
Section~\ref{sec:method} describes the formalism and computational
framework utilized.
We compute consistent, constraint-satisfying equilibrium
configurations with magnetic field strengths up to
${\sim}10^{17}$~G and analyze their global properties in Sec.~\ref{sec:sub:ini_conf}.
These configurations are then evolved using
general-relativistic magnetohydrodynamics in $(3+1)$ dimensions to
assess their stability and nonlinear dynamics. These results are
presented in Sec.~\ref{sec:sub:dyn}.
Finally, in conclusion we discuss in Sec.~\ref{sec:con} the implications of our
results for astrophysical pulsars and BNSM remnants.

Throughout the paper, we adopt geometric units $c=G=1$ with $\Mo=1$,
unless the units are explicitly indicated.


\section{Method}
\label{sec:method}


\subsection{Initial data}

Equilibrium configurations of magnetized neutron stars are constructed
by solving the Einstein constraint equations coupled to the equations
of ideal general-relativistic magnetohydrodynamics (GRMHD) under the assumptions of stationarity
and axisymmetry. We follow the approach developed in
Refs.~\cite{Bucciantini:2010ax,Bucciantini:2014zca,Pili:2014npa,Pili:2014zna,Pili:2017yxd},
in which the initial data problem is formulated using the conformal transverse-traceless 
decomposition.

In this framework, the spatial metric is assumed to be conformally
flat, $\gamma_{ij} = \psi^4 f_{ij}$, 
where $\psi$ is the conformal factor and $f_{ij}$ is the flat
three-metric. Maximal slicing is imposed, $K = 0$, which simplifies the
Einstein constraints into a set of coupled elliptic equations for the
metric potentials. Under these assumptions, the system reduces to a
well-posed elliptic problem with locally unique solutions
\cite{Cordero-Carrion:2008grk}.

For the matter sector, we assume a perfectly conducting fluid in ideal
GRMHD. The stress-energy tensor is given by
\begin{equation}
T^{\mu\nu} = (\rho h + b^2)u^\mu u^\nu +
\left(p + \frac{b^2}{2}\right)g^{\mu\nu} - b^\mu b^\nu,
\end{equation}
where $\rho$ is the rest-mass density, $p$ the pressure,
$u^\mu$ the fluid four-velocity, and $b^\mu$ the magnetic field
measured in the comoving frame. The specific enthalpy is
\begin{equation}
h = 1 + \epsilon + \frac{p}{\rho},
\end{equation}
where $\epsilon$ is the specific internal energy, and 
$b^2 \equiv b^\mu b_\mu$
is the magnetic-field strength measured in the comoving frame.

Stationarity and axisymmetry imply that the relativistic Euler
equation can be integrated to yield a Bernoulli-type first integral.
At the same time, the solenoidal constraint on the magnetic field,
$\partial_iB^i = 0$, allows one to introduce a magnetic
vector potential $A_\phi$ such that the poloidal magnetic field is expressed in
terms of its gradients. The equilibrium equations then reduce to a
Grad--Shafranov equation for $A_\phi$, coupled to the metric equations.

The magnetic field configuration is specified through two free
functions: the magnetization function $\mathcal{M}(A_\phi)$ and the
current function $\mathcal{I}(A_\phi)$, which are constant on magnetic
surfaces (see Eqs.~(32)--(33) in Ref.~\cite{Pili:2014npa}). In this
work, we restrict attention to purely poloidal magnetic field
configurations by setting $\mathcal{I}(A_\phi)=0$. The magnetization
function is chosen as
\begin{equation}
\mathcal{M}(A_\phi) = k_{\rm pol} \left(A_\phi+\frac{1}{2}\xi A_\phi\right),
\end{equation}
where $k_{\rm pol}\approx 0.008-0.06$ is varied to control the magnetic field
strength and $\xi$ is the nonlinear poloidal term in the current
distribution, which is set to zero in our simulations. This choice ensures regularity at the stellar center and
confinement of the field within the star.

The resulting system of elliptic equations is solved using the
publicly available \texttt{XNS} code~\cite{Bucciantini:2010ax,Bucciantini:2014zca,Pili:2014npa,Pili:2014zna,Pili:2017yxd} which is specifically designed
to compute equilibrium configurations of rotating and magnetized neutron
stars in general relativity.
We set \texttt{XNS} spherical with grid resolutions $(\Delta r,\Delta\theta)=(111\text{ m}, 1.7^\circ)$ and extending to radius $752.6$~km.
  Numerical experiments indicated that the setup is
  sufficient to accurately resolve both the
stellar interior and the magnetic field structure.

Neutron star matter is described by the finite-temperature SFHo
equation of state \cite{Hempel:2009mc,Hempel:2011mk,Steiner:2012rk},
which is consistent with current nuclear physics constraints and
astrophysical observations. Initial configurations are constructed assuming a
low, uniform temperature $T \sim 0.1~\mathrm{MeV}$, representing cold in $\beta$-equilibrium matter. Under these conditions, thermal contributions to the pressure are
subdominant, and the equilibrium structure is primarily determined by
the cold EOS component.


\subsection{(3+1)D Evolutions}

The dynamical evolution of the equilibrium configurations is performed
by solving the Einstein equations coupled to GRMHD in a $(3+1)$
decomposition of spacetime.
We adopt the Z4c free-evolution scheme of Einstein equations~\cite{Bernuzzi:2009ex, Ruiz:2010qj, Weyhausen:2011cg,Hilditch:2012fp}, which is 
designed to control the growth of constraint
violations through their continuum propagation properties and damping
terms. Gauge conditions are specified via the moving puncture gauge.

The GRMHD evolution is formulated in Eulerian 3+1 conservative form by evolving the
densitized Eulerian variables,
\begin{equation}
\mathbf{q} = \sqrt{\gamma}
\left(D, S_i, \tau, B^i \right),
\end{equation}
where $D$ is the conserved rest-mass density, $S_i$ the momentum
density, $\tau$ the internal energy density, and $B^i$ the magnetic
field. The system can be written as a set of 
flux-conservative equations,
\begin{equation}
\partial_t \mathbf{q} + \partial_j \mathbf{F}^j = \mathbf{s},
\end{equation}
where $\mathbf{F}^j$ are the fluxes and $\mathbf{s}$ the gravity
source terms, e.g.~\cite{Cook:2023bag}.
The magnetic field evolution satisfies the solenoidal constraint
$\partial_iB^i = 0$, which is enforced using a
constraint-transport scheme that preserves this condition to machine
precision (see below).
Matter is described during the evolution by the full finite-temperature
tabulated SFHo EOS.

All simulations are performed using the publicly available
\texttt{GR-Athena++} code \cite{Daszuta:2021ecf,Cook:2023bag,
  Daszuta:2024chz,Daszuta:2026szb}.
The numerical infrastructure employs high-resolution shock-capturing methods, using the local Lax–Friedrichs (LLF) approximate Riemann solver and piecewise parabolic method (PPM) reconstruction for the fluid variables. The magnetic field is evolved using the staggered-mesh constrained-transport scheme of~\citet{White:2015omx}, which maintains the solenoidal
constraint to machine precision.

The computational domain consists of a three-dimensional Cartesian
grid with no imposed symmetries. We employ a static mesh refinement
with nested grids centered on the star to adequately resolve
the high-density core while maintaining a large outer domain. The
domain extends to several stellar radii to minimize boundary
effects.

The computational domain consists of seven refinement levels. The
coarsest level covers the region $[-377.5,377.5]^3$ km$^3$ with $128$ grid
points per direction,
while successive refinement levels (starting at $[-94.37,94.37]^3$
km$^3$) halve the spatial extent, reaching a finest (7th) level spanning
$[-1.475,1.475]^3$ km$^3$ with a linear spacing of $\Delta x=46.08$~m in the finest
level and a spacing of $\Delta x=368.6$~m (4th level) always covering the whole
star. To assess the robustness of our results, we perform convergence
tests by decreasing the base resolution to $64$ points on the
coarsest grid, confirming that the main dynamical features and
collapse times are stable within numerical uncertainties.

To handle low-density regions outside the star, an artificial
atmosphere is introduced. The atmosphere is characterized by a floor
density $\rho_{\rm atm}\approx 6.2\times10^{7}$ g/cm$^3$, chosen
to be sufficiently small compared to the stellar values so as not to
affect the dynamics while maintaining numerical stability (see
Appendix~\ref{app:convergence} for convergence tests).

Black hole formation is monitored through the use of an apparent
horizon finder, which identifies marginally outer trapped surfaces during the
evolution. Once an apparent horizon is detected, properties such as
the horizon mass and spin are computed from the area law and
Christodoulou formula~\cite{Christodoulou:1970wf}.


\section{Results}
\label{sec:res}


\subsection{Initial configurations}
\label{sec:sub:ini_conf}

\begin{figure*}[t]
    \centering
    \includegraphics[width=\textwidth]{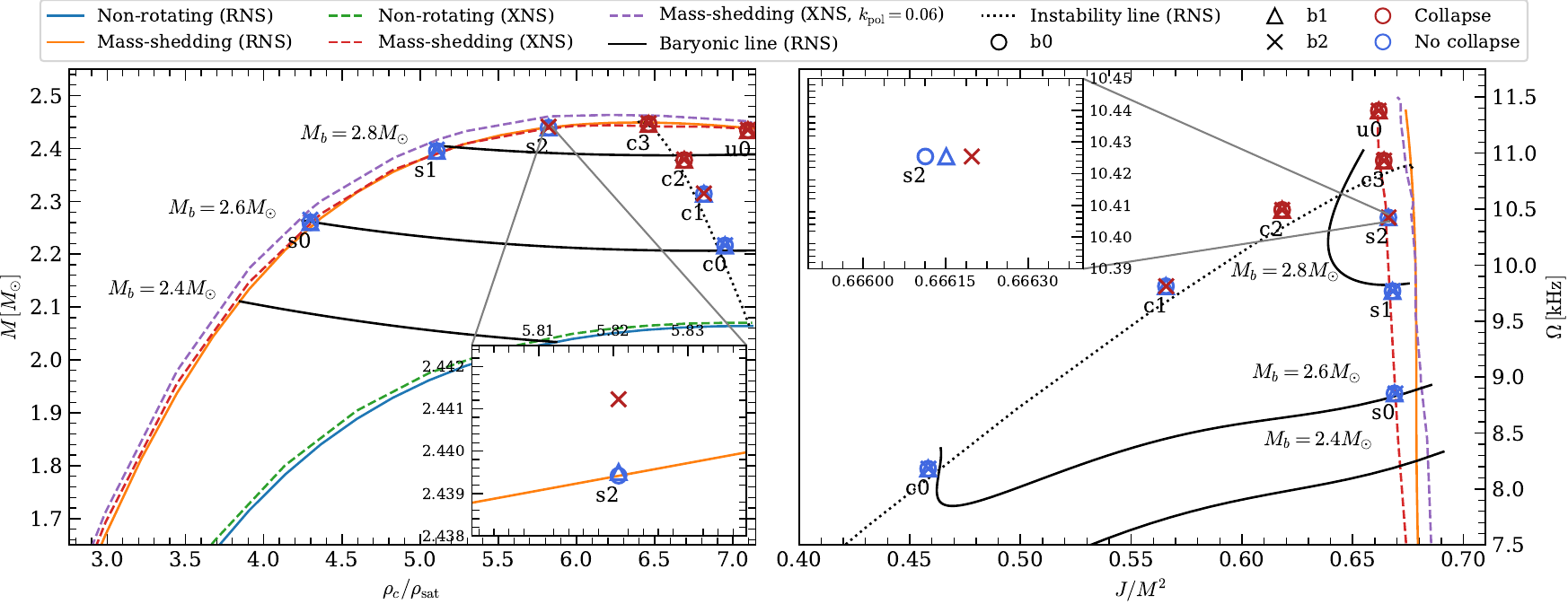}
    \caption{Initial data sequences (solid lines) and chosen configurations for dynamical evolutions (markers). Left: Gravitational mass $M$ and central density $\rho_c$ scaled by $\rho_{\rm sat}=2.7\times 10^{14}$ g/cm$^3$. Right: Angular velocity $\Omega$ and mass-rescaled angular momentum $J/M^2$ in geometric units ($c=G=\Mo=1$). Black lines indicate sequences at fixed baryon mass $M_b=2.4, 2.6, 2.8\Mo$, as indicated by the plot annotation. SMNSs correspond to equilibria in the regions above the the line $M_b\gtrsim2.4\Mo$. The black dotted line indicates the stability limit to gravitational collapse (for non-magnetized sequences). Non-magnetized configurations are obtained with both the \texttt{XNS} and \texttt{RNS} code. The purple line represents mass-shedding configurations with magnetic fields of order $B_{\rm max}\sim 1.5-2.5\times 10^{17}$ G. Configurations with same text label (c1, c2, etc.) correspond to the same values of $(\rho_c,\Omega)$. Each marker shape corresponds to a non, weakly or strongly magnetized configuration. Blue and red markers refer to configurations that are found dynamically stable and collapsing respectively.}
    \label{fig:XNS:data}
\end{figure*}

We construct sequences of SMNSs up to the mass-shedding
limit, where the stellar surface reaches Keplerian velocity at the
equator. The global properties of the equilibria sequences are shown
Fig.~\ref{fig:XNS:data}.
The left-side panel shows the gravitational mass versus the central
rest-mass density, normalized by the nuclear saturation density; the
right-side panel shows the angular frequency versus the normalized
angular momentum. We also show sequences of constant baryon mass as
black lines.

Initial data for non-magnetized equilibria are also computed with the 
\texttt{RNS} code~\cite{Stergioulas:1994ea}, which
does not assume a conformally flat metric. 
We find excellent agreement between the two codes, even for the most extreme cases, consistent with previous comparisons~\cite{Bucciantini:2010ax}.

Figure~\ref{fig:XNS:data} also shows the configurations
dynamically evolved and described in
Sec.~\ref{sec:sub:dyn}. Configurations s0-s2 are stable SMNSs masses
close to the mass-shedding limit with progressively larger masses. 
Configurations c0-c3 are SMNSs close to the
threshold for gravitational collapse; small perturbations during the
evolution can trigger collapse on dynamical timescales.
Configurations u0 are unstable SMNSs close to the mass-shedding limit.
We stress that configurations (s0, s1, s2, c3, u0) are chosen to lie slightly
below the mass-shedding threshold, typically by $\sim 1$--$3\%$ in angular
frequency and mass. 
Each configuration is characterized by three primary parameters: the
central rest-mass density $\rho_c$, the angular frequency $\Omega$,
and the magnetic field strength. For each pair $(\rho_c,\Omega)$, we
construct three configurations corresponding to different magnetic field intensities:

b0)~a non-magnetized configuration ($B=0$);

b1)~a ``weakly magnetized'' configuration ($B_{\rm max}\approx
10^{16}\,\mathrm{G}$) modeling magnetar strengths;

b2)~a strongly magnetized configuration ($B_{\rm max}\approx
10^{17}\,\mathrm{G}$) modeling BNSM remnant strengths.

We consider eight base stellar models, each evolved in three magnetic configurations: non-magnetized (b0), weakly magnetized (b1), and strongly magnetized (b2). This gives a total of 24 initial configurations.
As a result, configurations cluster in groups
with nearly identical hydrodynamical properties but different magnetic
energies, allowing us to isolate the dynamical impact of magnetic
fields.


\begin{figure}[t]
    \centering
    \includegraphics[width=.49\textwidth]{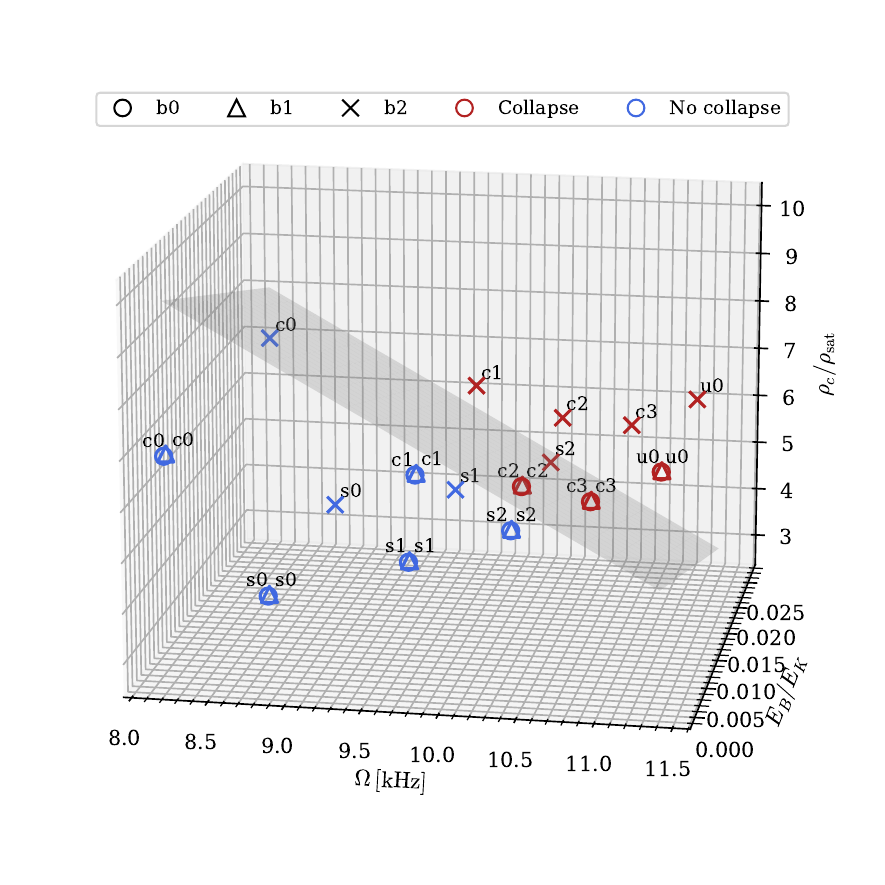}
    \caption{Central density ($z$-axis) as a function of $E_B/E_K$ ($y$-axis) and $\Omega$ ($x$-axis) for the configurations considered in this study. Blue and red markers refer to configurations that are found dynamically stable and collapsing respectively. The maximum-margin separating hyperplane has normal vector $\vec n=(3.6128, 161.4009, 2.2382)$ and $z$-intercept being $b=-51.8524$.} 
    \label{fig:Hyperplane3D}
\end{figure}

The inclusion of magnetic fields modifies the equilibrium sequences
and increases the maximum mass achievable at the mass-shedding limit
compared to non-magnetized configurations, see
Fig.~\ref{fig:XNS:data}. This is primarily due to the additional
magnetic stresses, which provide extra support against gravity and
allow equilibrium configurations with smaller equatorial radii at
fixed central density and rotation rate. In rotating stars,
centrifugal support dominates in the outer layers, while magnetic
stresses are more centrally concentrated and act to stabilize the
inner regions of the star. As a result, magnetized equilibrium
sequences typically exhibit slightly reduced radii and an increase in
the maximum supported mass relative to the purely hydrodynamic
case~\cite{Bocquet:1995je}. In our configurations, this effect is
modest but systematic: strongly magnetized configurations show (at most) a
0.2\% increase in the maximum mass and a corresponding 1\% 
reduction in equatorial radius. The mass-shedding for the b2 sequence,
in particular, achieves a maximum mass $M\simeq2.46\Mo$ for a
central density $\rho_{\rm max}\simeq1.67\times 10^{15}\,\gccm$ and angular
frequency $\Omega_{\rm max}\simeq11.50$~Hz. For comparison, the b0
sequence has 
$M\simeq2.44\Mo$ at $\rho_{\rm max}\simeq1.68\times 10^{15}\, \gccm$ and 
$\Omega_{\rm max}\simeq11.38$~Hz.

The stellar shape is only weakly deformed with respect to the non-magnetized case. Inspection of the equilibrium models shows nearly identical density isocontours. This is quantified by the superellipsoid index $n_s$, defined in Eq.~(56) of Ref.~\cite{Pili:2017yxd}, where $n_s=2$ corresponds to a perfect ellipsoid. Weakly magnetized and non-magnetized configurations have $n_s\approx1.6-1.9$, while strongly magnetized configurations differ by only a $\sim2\%$ increase from each respective index, indicating that the magnetic field induces only a very small difference with respect to the non-magnetized stellar shape for the magnetic fields considered in this work.

\begin{table*}
\centering
\caption{Main parameters of the ab-initio numerical relativity initial
  configurations used in this work. All masses are given in units of
$\Mo$. Configurations are generated with \texttt{XNS}. We set $\xi=0$ and
$k_{\rm pol}\approx 0.008-0.06$ such that we get $B_{\rm max}\sim 2-3.3\times10^{16}$ G
  for the intermediate magnetic field and $B_{\rm max}\sim
  1.5-2.5\times 10^{17}$ G for the high magnetic field. The columns
  ``B.N.'' and ``M.L.'' indicate the base name and the magnetic label
	of each configuration. The magnetic ($E_B$) and kinetic ($E_K$) energies are show as a ratio with respect to the gravitational binding energy ($\mathcal{W}$).}
\begin{tabular}{|c|c|c|c|c|c|c|c|c|c|c|c|c|}
\hline
B.N. & $\rho_c$[$10^{15}$ g/cm$^3$] & $\Omega$[kHz] & $P$[ms] & M.L. & $M$ & $M_b$ &  $B_{\text{max}}$[G] & $E_B/\mathcal{W}$& $E_K/\mathcal{W}$ & $J$ [$M_\odot^2$] & $R_{\text{e}}$[km] & $R_{\text{p}}/R_{\text{e}}$ \\
\hline
\multirow{3}{*}{c0} & \multirow{3}{*}{1.876} & \multirow{3}{*}{8.181} & \multirow{3}{*}{0.768} & b0 & 2.216 & 2.599 & 0 & 0 & $5.663\times 10^{-2}$ & 2.251 & 7.143 & 0.8295 \\
\cline{5-13}
 &  &  &  & b1 & 2.216 & 2.599 & $3.054\times 10^{16}$ & $2.688\times 10^{-5}$ & $5.663\times 10^{-2}$ & 2.251 & 7.143 & 0.8295 \\
\cline{5-13}
 &  &  &  & b2 & 2.217 & 2.599 & $2.292\times 10^{17}$ & $1.514\times 10^{-3}$ & $5.659\times 10^{-2}$ & 2.253 & 7.143 & 0.8295 \\
\hline
\multirow{3}{*}{c1} & \multirow{3}{*}{1.840} & \multirow{3}{*}{9.812} & \multirow{3}{*}{0.640} & b0 & 2.314 & 2.707 & 0 & 0 & $8.705\times 10^{-2}$ & 3.029 & 7.475 & 0.7481 \\
\cline{5-13}
 &  &  &  & b1 & 2.314 & 2.707 & $3.095\times 10^{16}$ & $2.974\times 10^{-5}$ & $8.704\times 10^{-2}$ & 3.029 & 7.475 & 0.7481 \\
\cline{5-13}
 &  &  &  & b2 & 2.315 & 2.707 & $2.323\times 10^{17}$ & $1.675\times 10^{-3}$ & $8.697\times 10^{-2}$ & 3.032 & 7.475 & 0.7481 \\
\hline
\multirow{3}{*}{c2} & \multirow{3}{*}{1.806} & \multirow{3}{*}{10.493} & \multirow{3}{*}{0.599} & b0 & 2.378 & 2.779 & 0 & 0 & $1.046\times 10^{-1}$ & 3.496 & 7.807 & 0.6879 \\
\cline{5-13}
 &  &  &  & b1 & 2.378 & 2.779 & $3.106\times 10^{16}$ & $3.173\times 10^{-5}$ & $1.046\times 10^{-1}$ & 3.496 & 7.807 & 0.6879 \\
\cline{5-13}
 &  &  &  & b2 & 2.379 & 2.779 & $2.137\times 10^{17}$ & $1.501\times 10^{-3}$ & $1.045\times 10^{-1}$ & 3.499 & 7.807 & 0.6879 \\
\hline
\multirow{3}{*}{c3} & \multirow{3}{*}{1.744} & \multirow{3}{*}{10.933} & \multirow{3}{*}{0.575} & b0 & 2.447 & 2.856 & 0 & 0 & $1.215\times 10^{-1}$ & 3.975 & 8.361 & 0.6291 \\
\cline{5-13}
 &  &  &  & b1 & 2.446 & 2.856 & $3.084\times 10^{16}$ & $3.414\times 10^{-5}$ & $1.214\times 10^{-1}$ & 3.974 & 8.361 & 0.6291 \\
\cline{5-13}
 &  &  &  & b2 & 2.448 & 2.857 & $2.315\times 10^{17}$ & $1.923\times 10^{-3}$ & $1.213\times 10^{-1}$ & 3.979 & 8.250 & 0.6376 \\
\hline
\multirow{3}{*}{s0} & \multirow{3}{*}{1.160} & \multirow{3}{*}{8.854} & \multirow{3}{*}{0.710} & b0 & 2.260 & 2.592 & 0 & 0 & $1.223\times 10^{-1}$ & 3.418 & 10.133 & 0.6284 \\
\cline{5-13}
 &  &  &  & b1 & 2.260 & 2.592 & $2.256\times 10^{16}$ & $3.922\times 10^{-5}$ & $1.223\times 10^{-1}$ & 3.418 & 10.133 & 0.6284 \\
\cline{5-13}
 &  &  &  & b2 & 2.265 & 2.598 & $1.696\times 10^{17}$ & $2.215\times 10^{-3}$ & $1.222\times 10^{-1}$ & 3.434 & 10.133 & 0.6284 \\
\hline
\multirow{3}{*}{s1} & \multirow{3}{*}{1.379} & \multirow{3}{*}{9.768} & \multirow{3}{*}{0.643} & b0 & 2.395 & 2.780 & 0 & 0 & $1.235\times 10^{-1}$ & 3.832 & 9.358 & 0.6213 \\
\cline{5-13}
 &  &  &  & b1 & 2.395 & 2.780 & $2.610\times 10^{16}$ & $3.762\times 10^{-5}$ & $1.235\times 10^{-1}$ & 3.833 & 9.358 & 0.6213 \\
\cline{5-13}
 &  &  &  & b2 & 2.398 & 2.783 & $1.797\times 10^{17}$ & $1.782\times 10^{-3}$ & $1.235\times 10^{-1}$ & 3.842 & 9.247 & 0.6287 \\
\hline
\multirow{3}{*}{s2} & \multirow{3}{*}{1.572} & \multirow{3}{*}{10.425} & \multirow{3}{*}{0.603} & b0 & 2.439 & 2.845 & 0 & 0 & $1.228\times 10^{-1}$ & 3.964 & 8.693 & 0.6306 \\
\cline{5-13}
 &  &  &  & b1 & 2.440 & 2.845 & $2.874\times 10^{16}$ & $3.581\times 10^{-5}$ & $1.228\times 10^{-1}$ & 3.964 & 8.693 & 0.6306 \\
\cline{5-13}
 &  &  &  & b2 & 2.441 & 2.847 & $1.978\times 10^{17}$ & $1.695\times 10^{-3}$ & $1.227\times 10^{-1}$ & 3.970 & 8.693 & 0.6306 \\
\hline
\multirow{3}{*}{u0} & \multirow{3}{*}{1.916} & \multirow{3}{*}{11.379} & \multirow{3}{*}{0.552} & b0 & 2.434 & 2.840 & 0 & 0 & $1.197\times 10^{-1}$ & 3.922 & 8.029 & 0.6276 \\
\cline{5-13}
 &  &  &  & b1 & 2.434 & 2.840 & $3.275\times 10^{16}$ & $3.250\times 10^{-5}$ & $1.197\times 10^{-1}$ & 3.922 & 8.029 & 0.6276 \\
\cline{5-13}
 &  &  &  & b2 & 2.435 & 2.840 & $2.458\times 10^{17}$ & $1.829\times 10^{-3}$ & $1.195\times 10^{-1}$ & 3.925 & 7.918 & 0.6364 \\
\hline
\end{tabular}
\label{tab:initial_configs}
\end{table*}

The main properties of all configurations are reported in
Table~\ref{tab:initial_configs}. Global quantities such as the
gravitational mass $M$ and baryonic mass $M_b$ vary only marginally
across the different b0-2 configurations. The small relative
differences reported in the table reflect the fact that magnetic
energy is subdominant with respect to the kinetic energy in all configurations, with
$E_B/E_K \lesssim 3\times10^{-2}$. Rotating configurations where magnetic and
kinetic energies are comparable have been explored in other studies
but considering lower rotation rates~\cite{Pinas:2025bpq}, \ie~they
do not consider SMNSs.

Reaching the regime in which magnetic and kinetic energies become
  comparable for our configurations would require field 
strengths $B \gtrsim 5\times10^{17}$--$10^{18}\,\mathrm{G}$~\cite{Pili:2017yxd}.
However, such high magnetic fields are not accessible
in our dynamical simulations due to numerical limitations.
In practice, we find that numerically stable GRMHD evolutions
with our setup 
are restricted to magnetic field strengths 
${\lesssim} 3\times10^{17}\,\mathrm{G}$.
This limit is set by the large magnetizations ($\propto B^2/\rho$)
that develop in the low-density atmosphere, which challenge the
primitive-variable recovery and other GRMHD algorithms; evolving
stronger fields would require either a substantially denser
atmosphere, which can affect the dynamics, or more specialized
treatments of magnetically dominated regions. 

Figure~\ref{fig:Hyperplane3D} shows a 3D representation of the central
density as a function of the ratio of magnetic to kinetic energy and
the angular velocity, $\Omega$, for the configurations listed in
Table~\ref{tab:initial_configs}. The figure also identifies which
configurations undergo gravitational collapse and which remain stable
during the subsequent dynamical evolutions discussed in later
sections. The collapsing and non-collapsing configurations occupy distinct
regions of the parameter space, while configurations close to the
instability threshold lie near the boundary between these two
populations. To better characterize this transition, we computed the
maximum-margin separating hyperplane using
\texttt{sklearn}~\cite{Pedregosa:2011ork}, which provides an
approximate 
empirical boundary between collapsing and non-collapsing
configurations. Within the explored parameter space, this classifier
suggests that, for a given central density, configurations with both
higher rotation rates and stronger magnetic fields are more prone to
collapse.
The results indicate that stability near the supramassive threshold is
governed by the combined influence of rotation, compactness, and
magnetic effects, rather than by any single global quantity. In the
following sections, we examine the dynamical evolution of these configurations
to identify the mechanisms that ultimately determine whether a
configuration collapses. 


\subsection{Dynamical evolutions}
\label{sec:sub:dyn}


\subsubsection{Overview of stability results}
\label{sec:sub:sub:stability}

\begin{figure*}[t]
    \centering
    \includegraphics[width=0.8\textwidth]{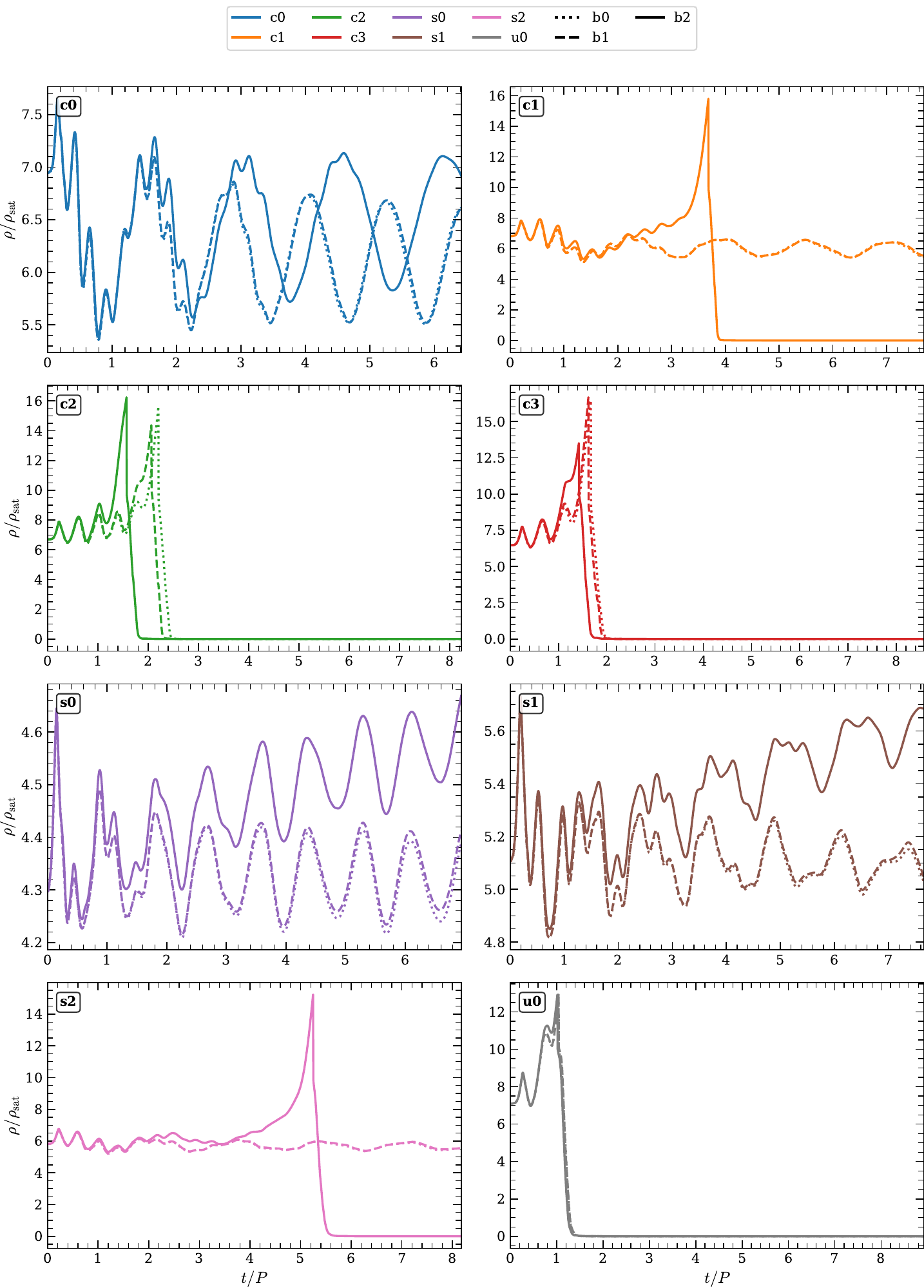}
    \caption{Evolution of the central density (rescaled by $\rho_{\rm sat} = 2.7\times 10^{14}\, \gccm$) for all SMNS configurations in Tab.~\ref{tab:initial_configs}.}
    \label{fig:rho_rat}
\end{figure*}

As a proxy for the dynamical stability of the initial data we consider
in Fig.~\ref{fig:rho_rat} the evolution of the maximum (central)
density for all configurations. In absence of magnetic fields, the
configurations belonging to the stable 
branch (s*b0) remain stable throughout the simulated time. The central
density exhibits small oscillations around the initial equilibrium value due to
numerical truncation error. The most extreme configurations beyond the
point of stability collapse on dynamical timescales; 
collapse is triggered by truncation errors that perturb the
unstable equilibria. Notably, resolution is sufficiently high so that
configurations c0 and c1, close to the line of stability, do not
collapse on the simulated timescales. Configuration u0b0
collapses after approximately one
rotation period ($t/P\sim1$), while configurations c2b0 and c3b0 collapse
after roughly two rotation periods ($t/P\sim2$).

The impact of magnetic fields on the stellar dynamics depends strongly
on both the initial compactness and the magnetic field
strength. For initially weaker magnetic fields (b1 series),
the central density evolution is qualitatively similar to the
non-magnetized case. Each b1 configurations either remains stable over
the simulated timescale or collapses at comparable timescale of 1-2 periods.
In contrast, strongly
magnetized configurations (b2 series, $B_{\rm max}\sim10^{17}\,\mathrm{G}$) show
significant deviations. Magnetic stresses efficiently redistribute
angular momentum within the star, eventually accelerating the onset of
gravitational collapse. Configuration s2b2, in particular, is close to marginal
stability (\cf~s2b0), but collapses due to the presence of
strong magnetic fields.
This suggests that magnetic stresses can act as a catalyst for instability
in marginally stable systems. 

The dynamics of the simulated configurations is 
essentially governed by the interplay of rotation and magnetic fields 
instabilities.
In the earlier stages of the evolution the (unstable) poloidal
magnetic field develops the Tayler instability and a mixed
poloidal-toroidal configuration. Differential rotation also develops,
similarly to what observed in \cite{Tsokaros:2021pkh,Pinas:2025bpq,Joshi:2026eci}.
The outer layers of the star initially expand around the equatorial
plane; at later times the entire rotational profile up to the inner
core is affected. During this rearrangement, magnetic winding and 
magnetorotational (MRI) instability start to operate and 
redistribute angular momentum on Alfv\'en timescales.
If the latter
timescale is shorter than the rotational timescale,
angular momentum transport is particularly efficient.
Consistently, we observed the development of
differential rotation in all strongly magnetized
configurations on timescales of $t/P \sim 2$.
As magnetic stresses extract angular momentum from the inner core, the
local angular velocity decreases and the centrifugal support is
(locally) progressively reduced. Marginally stable configurations are
particularly sensitive to this effect: even a moderate reduction in
central rotation can increase the core density sufficiently to trigger
collapse.

In the following we discuss magnetic field evolution, differential
rotation and remnants in some detail.


\subsubsection{Magnetic field evolution}

\begin{figure}[t]
    \centering
    \includegraphics[width=.49\textwidth]{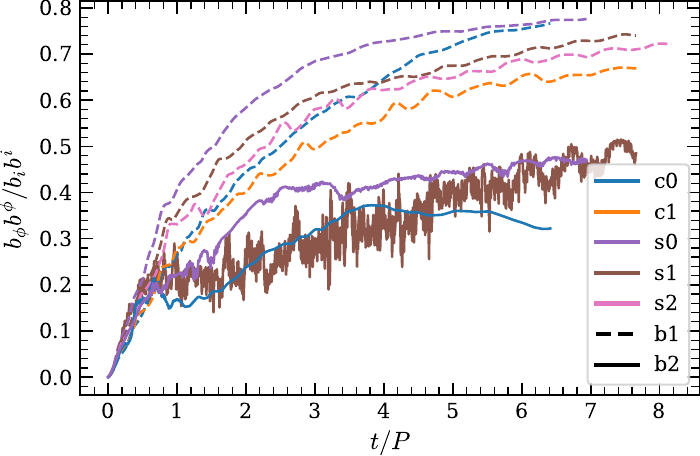}
    \caption{Evolution of the ratio of toroidal and total magnetic field for the stable magnetic configurations.}
    \label{fig:hst_toroidal}
\end{figure}

\begin{figure}[t]
    \centering
    \includegraphics[width=.49\textwidth]{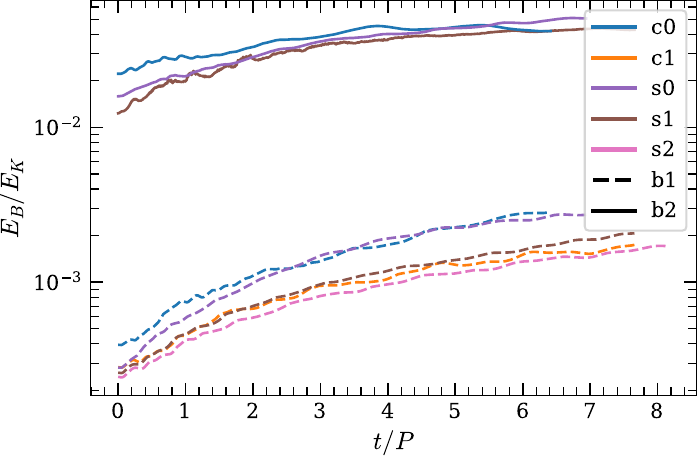}
    \caption{Evolution of the ratio of magnetic energy and kinetic energy
      for the stable magnetic configurations.}  
    \label{fig:hst}
\end{figure}

We define the Alfv\'en timescale as
\begin{equation}
	\tau_A \sim \frac{2R}{\langle v_A\rangle},
	\label{eq:alfven}
\end{equation}
where $R$ is a characteristic stellar radius (defined as the
equatorial radius that contains densities above $10^{13}$ g/cm$^3$)
and $\langle v_A\rangle$ is the average Alfv\'en velocity within the
interior of the star. The Alfv\'en velocity is defined as 
\begin{equation}
	v_A^2 = \frac{b^2}{b^2+\rho h}.
	\label{eq:va}
\end{equation}
The magnetic field configurations considered here are initially purely
poloidal. There are two mechanisms by which this system can generate a
toroidal field component. First, through the Tayler instability,
which acts on purely poloidal fields on Alfv\'en timescales
\cite{Tayler:1957a}. This disrupts the poloidal field structure
through an $m=0$ (varicose) and $m=1$ (kink) instability, generating a
toroidal component.  Second, through the winding driven by
differential rotation. Assuming an initially axisymmetric profile, the
growth rate of the toroidal component through winding is proportional
to the magnitude of differential rotation present in the system,
\be
\p_t B^\phi  = B^a\p_a \Omega
\ee
where $a = (r,\theta)$ are the poloidal components, and is therefore linear in time.

For the initially rigidly rotation configuration considered here, the only
mechanism initially available to generate a toroidal field then is the
Tayler instability. As the toroidal field develops, the
Maxwell stress term $T^{r\phi} \sim b^rb^\phi$ grows, and acts to
transport angular momentum outwards in the radial direction therefore 
developing a differentially rotating profile. We discuss the
development of this differential rotation in more detail in
Sec. \ref{sec:sub:sub:diffrot}.
As a consequence winding sets in and the toroidal field grows further. 

Figure \ref{fig:hst_toroidal} shows the growth of the toroidal
component over time. Over the initial period, the growth of the
toroidal field is approximately linear in time for all configurations,
and therefore compatible with a winding driven evolution. After the
first rotational period, two behaviours emerge for different initial
magnetic field strengths. The more weakly magnetised b1 configurations
continue to grow their toroidal magnetic field to values between
$60-80\%$ of the total magnetic field, as the winding process
continues. For the more strongly magnetised configurations b2, the
growth of the toroidal field gradually saturates to values between
$25-45\%$. This behaviour is 
also reflected in the overall evolution of magnetic energy relative to
rotational energy in Fig. \ref{fig:hst}. Weakly magnetized
configurations show a consistent amplification, while strongly
magnetized configurations approach a quasi-stationary state more
rapidly (if they do not collapse). 

The s1b2 (brown) curve in Fig.~\ref{fig:hst_toroidal} exhibits rapid fluctuations, reflecting the dynamical interplay between the strong magnetic field and rapid rotation in this configuration. These fluctuations suggest that the simulation is approaching the limits of the parameter space that can be reliably explored with our current GRMHD setup. Although we have verified that the adopted resolution is sufficient to resolve the relevant dynamics, we caution that configurations with even stronger magnetic fields and/or faster rotation may require additional numerical investigation to ensure the robustness of the results. 

As the coherent toroidal field grows in the remnant through winding,
it becomes again unstable to the Tayler instability. The latter 
disrupts the toroidal field, and convert it back into poloidal
field. This process arrests the development of the strong toroidal
field in the star. The Tayler instability acts on timescales of the
Alfv\'en time $\tau_A$, while the growth-rate of the toroidal field
depends on the degree of differential rotation. 

\begin{figure*}[t]
	\centering
	\includegraphics[width=0.9\textwidth]{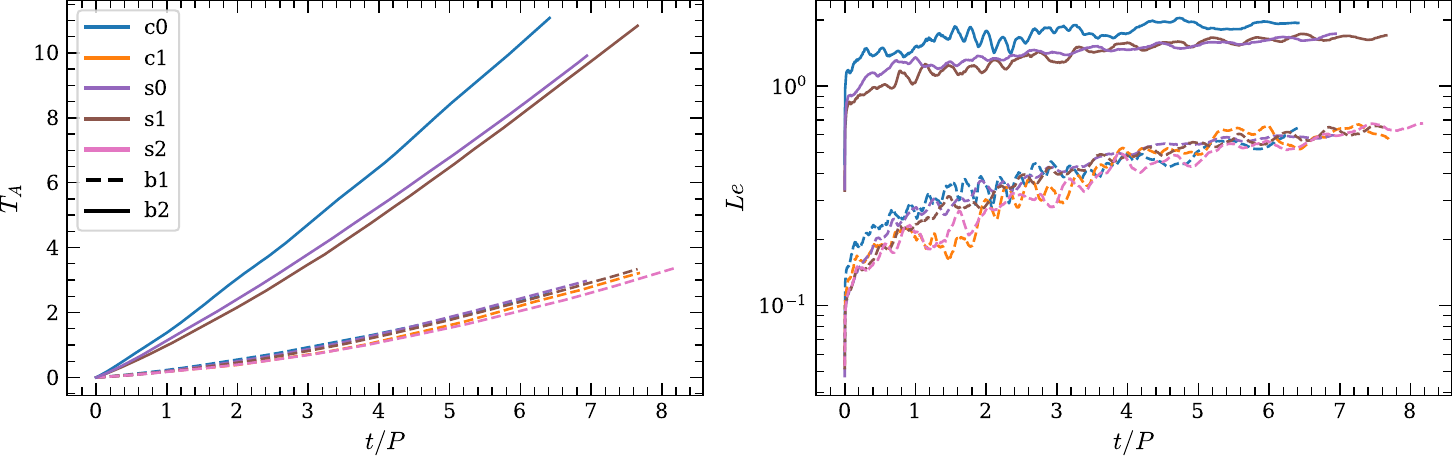}
	\caption{Evolution of the Alfv\'en crossing
		time and Lehnert number ($\Le$) for all non-collapsing configurations.}
	\label{fig:alf}
\end{figure*}

We measure the ratio of these timescales with the Lehnert number, $\Le
= P/\tau_A$. In Fig. \ref{fig:alf} we demonstrate the Lehnert number
for these different configurations, and see that the b2 configurations
have $\Le \gtrsim 1$, while the b1 configurations have $\Le < 1$. We
also consider the cumulative 
Alfv\'en crossing time~\cite{Sur:2021awe,Pinas:2025bpq} for all configurations, defined as  
\begin{equation}
	T_A = \int_0^t \frac{dt'}{\tau_A(t')},
	\label{eq:alfv,crossing}
\end{equation}
which measures the number of effective Alfv\'en times elapsed during
the evolution.
Weakly magnetized configurations
reach values of $T_A \gtrsim 3$ over the simulated time, while
strongly magnetized configurations reach $T_A \sim 11$. 
These correspond to Alfv\'en times $\tau_A \sim 0.2\mathrm{ms}$ ($\tau_A \sim
0.6\mathrm{ms}$) for the strongly (weakly) magnetized configurations.
 
In the b1 configurations, the Alfv\'en timescale is much longer than
the rotational timescales, so the Tayler instability cannot
efficiently disrupt the toroidal field as it develops through
winding, and by the end of the simulation, a strong toroidal field
 has developed as seen in
 e.g. \cite{Kiuchi:2008ss,Skoutnev:2024vlm}. In contrast, in the b2
 configurations the Tayler instability acts faster than the rotation
 time scale, and disrupts the toroidal field as fast as it can be
 generated through winding. The development of the toroidal field
 therefore stalls, and a weaker field develops.  
 
 \begin{figure*}[t]
   \centering
   \includegraphics[width=0.9\textwidth]{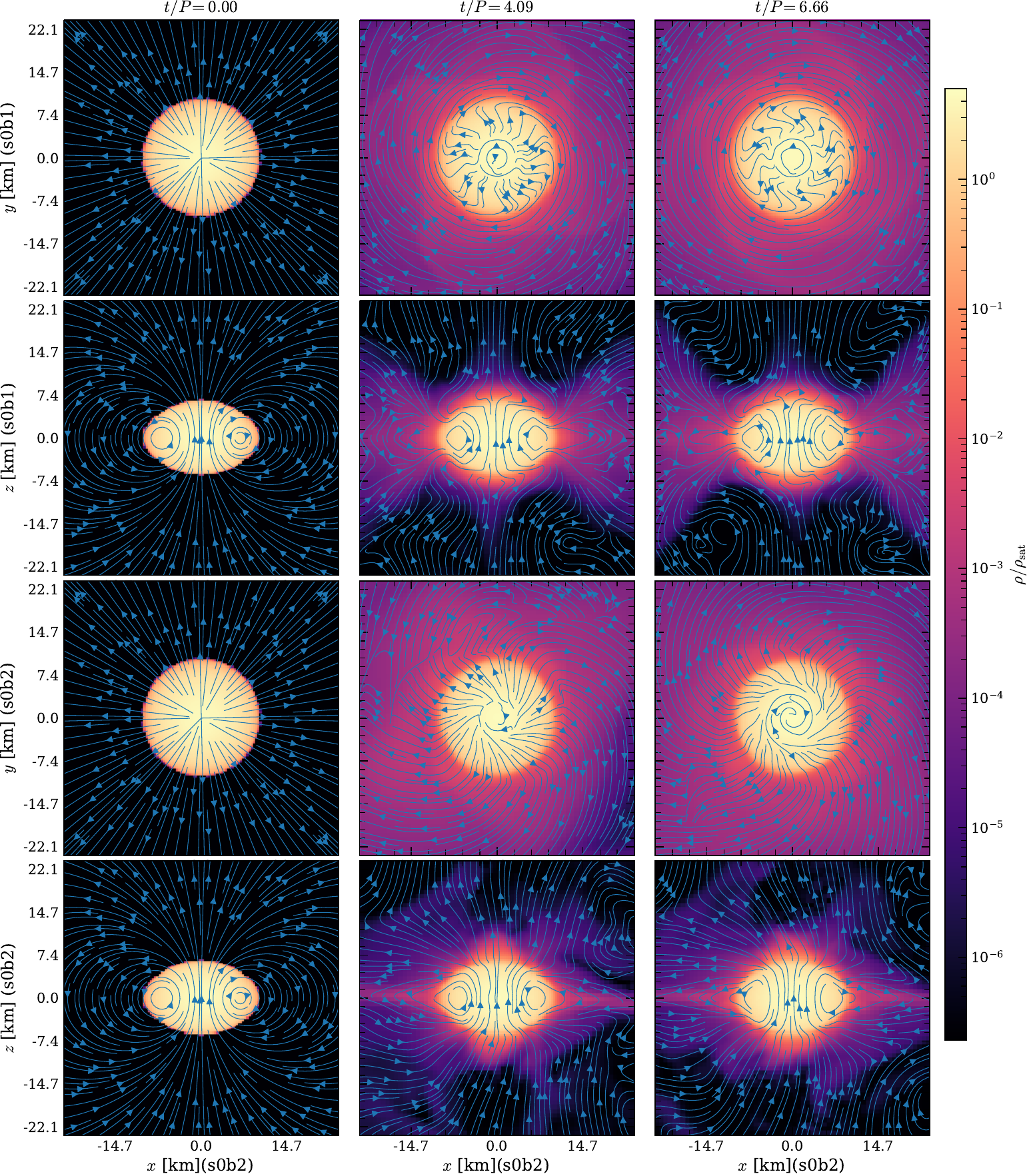}
   \caption{Snapshots of the density and magnetic field lines
     in the equatorial plane (top two rows) and the $xz$
     plane (bottom two rows) for the s0b1 and s0b2
     configurations. The three columns correspond to three specific evolution times.} 
   \label{fig:rho_xy_xz}
 \end{figure*}
  
Figure \ref{fig:rho_xy_xz} illustrates the development of these
processes in spatial slices of s0b1 and s0b2.
During the early evolution, $t\sim4.09P$, the initially poloidal
magnetic field is rapidly restructured as a toroidal  
component develops, leading to the formation of a mixed
poloidal--toroidal magnetic configuration. We note that a clear
difference is already visible between the b1 and b2
configurations. In the s0b1 configuration, a relatively compact
layer of low-density matter surrounds the star, with a more
pronounced vertical extent. In contrast, the s0b2 configuration
develops a lower and more radially extended layer of low-density
matter, particularly in the equatorial plane. As discussed in more
detail in later sections, these differences in the spatial
distribution are related to the distinct mass-ejection dynamics of
the b1 and b2 configurations.

Differences in the magnetic field structure are also already visible
at these early times. Configuration s0b1 has an ordered toroidal field both
inside the star and in the low-density envelope, as seen in the $xy$
plane. In the $xz$ plane, the poloidal field inside the star is disrupted. In
contrast, the s0b2 configuration has inner field closer to the
initial poloidal configuration, with a stronger radial component in
the $xy$ plane.

By $t\sim4.09P$ the growth of the toroidal field has been inhibited by
the Tayler instability, as indicated by the field
line structures. For the s0b1 configurations, the field
lines within the star in the $xz$ plane are poloidal, and close to
vertical in the $z$ direction, close to the $z$ axis. In contrast, for
the s0b2 configuration, the field lines inside the star are bent, the
sign of the Tayler instability displacing the field in the $xy$
plane, suppressing the toroidal field growth.

At late times, $t=6.66P$, the s0b2 configuration
shows a more turbulent magnetic field structure created in the
exterior. Vortex-like structures visible in the $xy$ plane create
counter-rotating magnetic field components. This is not observed in
s0b1, which shows a coherent toroidal field.

The development of the Tayler instability is further analysed by performing a decomposition of the magnetic field into azimuthal modes, following e.g. ~\cite{Lasky:2011un,Cook:2025zzy}. 
Specifically, we consider the projections
\begin{equation}
  C_m(t)=\frac{1}{\pi R_0^2}\int_0^{R_0}\, r\, dr\,
  \int_0^{2\pi}d\phi\, B_{\phi}(r,\theta=\pi/2)e^{im\phi}, 
\label{eq:fourier}
\end{equation}
which provides a global measure of the non-axisymmetric magnetic structure.

\begin{figure}[t]
  \centering
  \includegraphics[width=.49\textwidth]{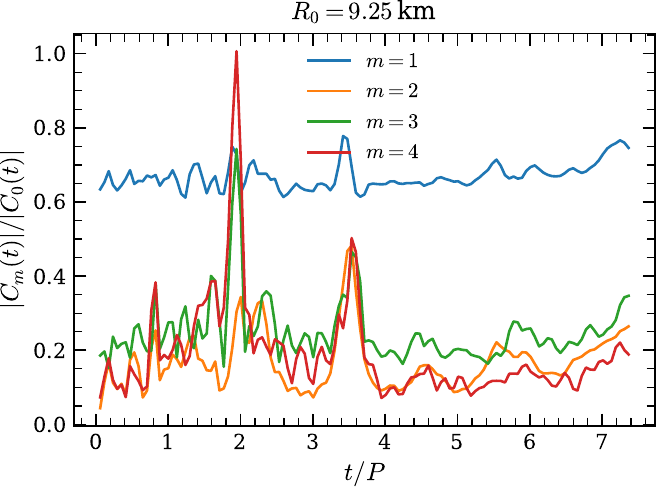}
  \caption{Evolution of the integrated Fourier modes (see
    Eq.~\eqref{eq:fourier}) for s0b2 at the initial radius (${\sim}9.25$~km). The magnitude represented is the
    magnitude of each mode normalized with the mode
    $m=0$.}
  \label{fig:fourier}
\end{figure}

Figure \ref{fig:fourier} shows the evolution of these modes for the
s0b2 case. Throughout the evolution we find that the $m=1$ mode is consistently $\sim 70\%$ of the $m=0$ mode magnitude, with higher modes
contributing up to 50\% of the $m=0$ field strength, with a sudden increase in $m=3$ and $m=4$ at $t/P\sim 2$. The magnitude of
the $m=1$ mode indicates the kink instability is acting to destabilise
the toroidal field and generate poloidal field components. In comparison, the weaker
magnetic field in s0b1 (not shown) is dominated by the $m=0$ mode. In
this case, the toroidal field grows without being
impacted by the Tayler instabilities.

\citet{Pinas:2025bpq,Joshi:2026eci} evolved rigidly rotating NSs with
initial poloidal magnetic fields similar to those considered here, but
with lower masses, \ie~they did not consider SMNSs. Their
magnetically dominated configurations ($\Le>1$) exhibit a clear decay
of the magnetic field associated with Tayler instabilities. By
contrast, our SMNS configurations rotate substantially faster,
remaining in a regime where rotation partially suppresses these
instabilities. Consequently, we observe an overall amplification of
the toroidal magnetic field and, even for the strongly magnetized b2
configurations, no magnetic energy decay on the simulated
timescales. While the $F_1B_2$ model of \citet{Joshi:2026eci} is the
closest analogue to our b2 configurations, its Alfv\'en timescale is
still approximately one order of magnitude larger, and their fastest
rotating models remain nearly a factor of two slower than our slowest
one. These differences place our simulations in a more
rotation-dominated regime, naturally explaining the opposite behavior
with respect to magnetic dissipation despite the development of
differential rotation in both studies. Nevertheless,
\citet{Joshi:2026eci} also report a transient growth of magnetic
energy in their $F_1B_1$ model and find that faster rotation enhances
the toroidal component. In particular, the toroidal energy in their
$F_2B_1$ model grows to approximately twice the poloidal energy by the
end of the simulation, while this ratio reaches roughly four in the
faster-rotating $F_1B_1$ model. This trend toward a stronger toroidal
component and a larger fraction of the magnetic energy being stored in
the toroidal field provides a possible connection between the
magnetic-energy decay observed in their generally slower-rotating
models and the sustained magnetic-field amplification found in our
more rapidly rotating SMNS configurations.


\subsubsection{Development of differential rotation}
\label{sec:sub:sub:diffrot}

\begin{figure*}[t]
    \centering
    \includegraphics[width=0.9\textwidth]{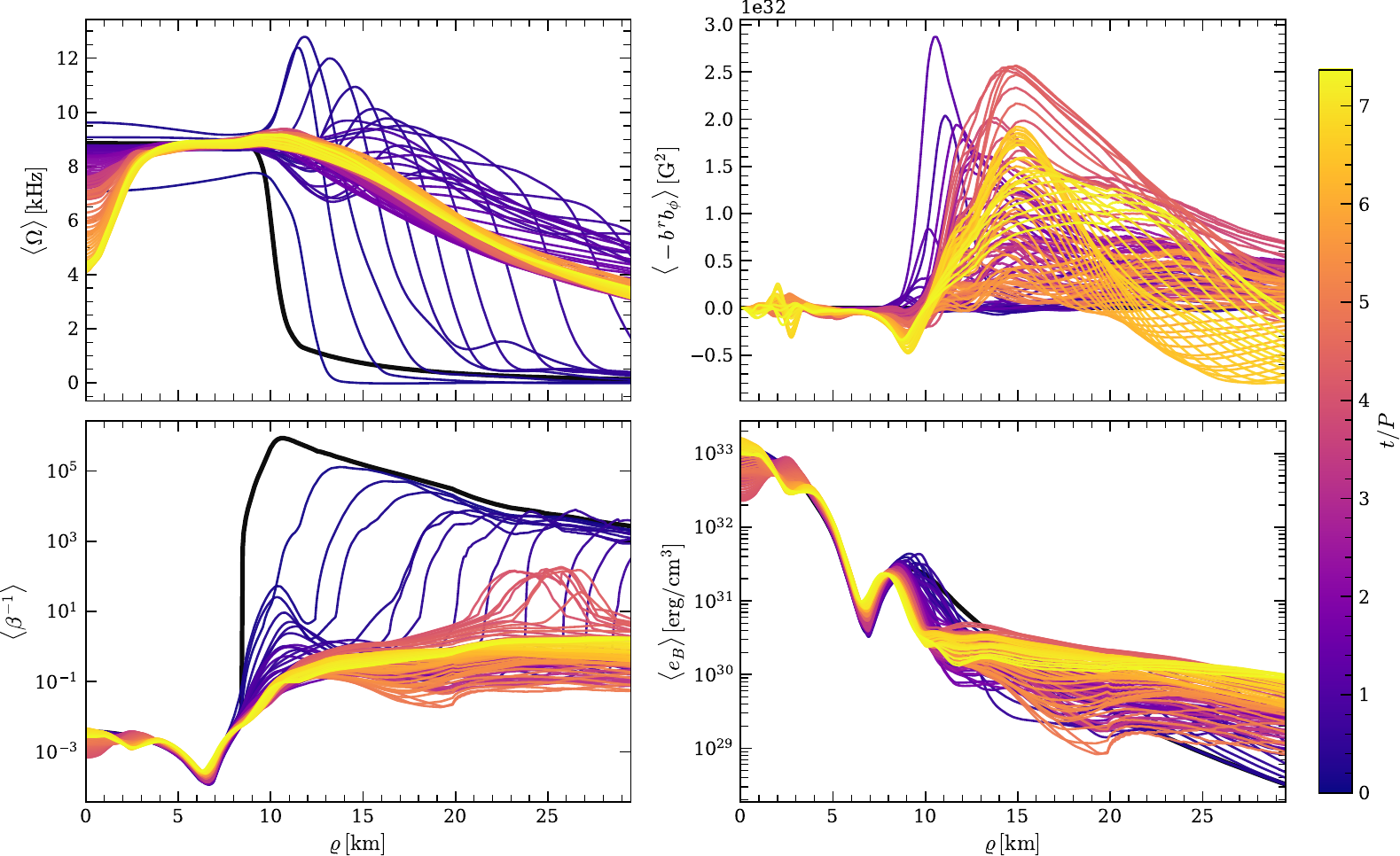}
    \caption{Time evolution of the azimuthally averaged angular
      velocity (top left panel), magnetic stress (top right panel),
      fraction of magnetic and fluid pressure (bottom left), and magnetic
      energy density (bottom right) profiles from s0b2 as a
      function of the cylindrical  
      radius $\varrho=\sqrt{x^2+y^2}$.}
    \label{fig:properties_4panel}
\end{figure*}

As discussed above, once a toroidal field develops through the Tayler instability, Maxwell stresses transport angular momentum radially outwards, leading to the growth of differential rotation.



The development of differential rotation is illustrated in
Fig.~\ref{fig:properties_4panel} (top left panel), which shows the azimuthally averaged spatial
profiles of the angular velocity and pressure ratio as function of the cylindrical radius
$\varrho=\sqrt{x^2+y^2}$ for configuration s0b2.
Different colors correspond to different times in the simulation.

The angular velocity profile clearly develops a
radially-dependent structure characterized by an outer layer extending
beyond the initial star radius and with $d\Omega/d\varrho<0$. This
region becomes quickly unstable to 
MRI which determines the final profile on timescales of a few
periods.
Magnetic stresses ($T^{r}_{\ \phi}$, top right panel) peak at
$\varrho\sim10$~km at early times, transport angular momentum
outwards, and progressively reduce over Alfv\'en timescales. 
We remark that the MRI quality factor in our simulations lies in the range
$Q \in [1,10]$, \cf~\cite{Gutierrez:2025gkx,Siegel:2013nrw}, with
characteristic wavelengths $\lambda \sim 1.47$ km. Therefore, such
instability may not be fully resolved in our simulations. 

We also observe that the inner core, $\varrho\lesssim 1$~km, spins down
relative to the initial angular velocity. Such regions rotate with
approximately $40-50\%$ less angular velocity at late times than at early
times, and slower than most of outer regions of the star.
The central region remains MRI stable throughout
the evolution, being characterized by $d\Omega/d\varrho>0$. 

The bottom left panel of Fig.~\ref{fig:properties_4panel} shows the 
ratio of the magnetic pressure to the plasma pressure (inverse of the
plasma-$\beta$ parameters), 
\begin{equation}
\beta^{-1}=\frac{p_{\rm mag}}{p}\propto \frac{B^2}{p}.
\end{equation}
The star's outer layers evolve from an initial magnetically
dominated state 
to a slightly pressure dominated at late times with
$\langle\beta^{-1}\rangle(\varrho\sim25{\rm km})\sim0.5$ and
$\langle\beta^{-1}\rangle(\varrho\sim10{\rm km})\sim0.1$.

The interior remains overall pressure dominated with the star becoming 
less compact along the evolution. However, the density in the inner core 
increases due to the reduced rotational support. The decrease in
$\beta^{-1}$ in the inner core is due to both an increase in 
central pressure and a decrease in magnetic pressure with respect
to the initial configuration.

The bottom right panel of Fig.~\ref{fig:properties_4panel} shows the
azimuthally averaged magnetic energy density, \(e_B=W\, b^2/2\), which
is related to the total magnetic energy \(E_B\) used in
Fig.~\ref{fig:hst} through
\begin{equation}
E_B = \int e_B \sqrt{\gamma} d^3x,
\end{equation}
where \(\gamma\) is the determinant of the spatial metric. Magnetic
energy is initially concentrated in the stellar interior and gradually
redistributes toward larger radii as the evolution proceeds. A local
maximum clearly tracks the peak of the angular velocity and magnetic
stresses. 

The configuration s0b2 well represents the behavior observed in all
strongly magnetized configurations b2. Configurations s0b2 and s1b2
appear to achieve a new, differentially rotating and dynamically
stable configuration. By contrast, the reduced pressure support in the
inner core triggers the collapse of s2b2.
However, weakly magnetized (b1) configurations do not show significant
development of differential rotation on the simulated timescale.

Differently from \citet{Tsokaros:2021pkh} we do not observe the
development of a counter-rotating inner core (as was also observed in
\citet{Pinas:2025bpq}). The polytropic configurations studied in the
first reference are 
characterized by an initially mixed (poloidal-toroidal) magnetic field
but are slower rotating than those considered here, \ie~not SMNSs.
Their configuration A6 is the only SMNS, marginally stable,
configuration simulated there, and its collapse behavior is indeed
very similar to our s2b2.


\subsubsection{Black hole properties}
\label{sec:sub:sub:bh}

\begin{table}[t]
\centering
\caption{Properties of remnant black holes from unstable
  configurations. The columns ``B.N.'' and ``M.L.'' indicate the base name and the magnetic label of each configuration.}
\begin{tabular}{|c|c|c|c|c|c|c|}
\hline
B.N. & M.L. & $t_{\text{col}}/P$ & $M_{\text{BH}}$ & $S_H/M_{\text{BH}}^2$ & $\Omega_{H}$[kHz] \\
\hline
\multirow{1}{*}{c1} & b2 & 3.653 & 2.282 & 0.551 & 13.356 \\
\hline
\multirow{3}{*}{c2} & b0 & 2.205 & 2.349 & 0.625 & 15.167 \\
\cline{2-6}
 & b1 & 2.065 & 2.349 & 0.624 & 15.136 \\
\cline{2-6}
 & b2 & 1.571 & 2.347 & 0.614 & 14.840 \\
\hline
\multirow{3}{*}{c3} & b0 & 1.663 & 2.416 & 0.670 & 16.159 \\
\cline{2-6}
 & b1 & 1.620 & 2.417 & 0.672 & 16.216 \\
\cline{2-6}
 & b2 & 1.423 & 2.400 & 0.668 & 16.214 \\
\hline
\multirow{1}{*}{s2} & b2 & 5.247 & 2.388 & 0.643 & 15.469 \\
\hline
\multirow{3}{*}{u0} & b0 & 1.044 & 2.387 & 0.686 & 16.887 \\
\cline{2-6}
 & b1 & 1.044 & 2.387 & 0.686 & 16.877 \\
\cline{2-6}
 & b2 & 1.017 & 2.379 & 0.667 & 16.293 \\
\hline
\end{tabular}
\label{tab:BHs}
\end{table}

\begin{figure}[t]
    \centering
    \includegraphics[width=.49\textwidth]{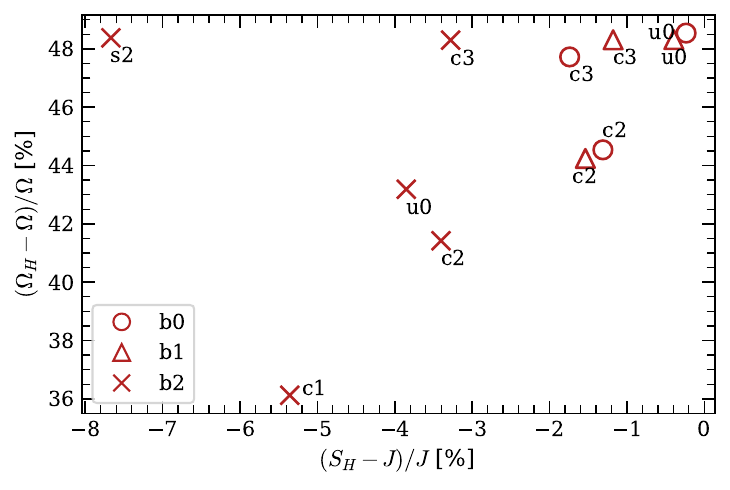}
    \caption{Relative change in the angular velocity and
      angular momentum between the initial stellar configuration and
      the final black hole for all collapsing configurations.}
    \label{fig:delta_om_jm2}
\end{figure}

The properties of the remnant black holes for collapsing configurations
are collected in Table~\ref{tab:BHs}.
The angular velocity of the black hole horizon, $\Omega_H$, is
computed from the mass and spin of the remnant using
\begin{equation}
\Omega_{\rm H}=\frac{S_{\rm H}}{2M_{\rm H}^2\left(M_{\rm H}+\sqrt{M_{\rm
      H}^2-(S_{\rm H}/M_{\rm H})^2}\right)} .
\end{equation}
We find that $\Omega_{\rm H}$ increases with both the compactness of the
progenitor and its rotational energy, reaching values above ${\sim}17$
kHz for the most extreme configurations. Comparing the initial stellar
angular velocities from Table~\ref{tab:initial_configs} with the
corresponding black hole values in Table~\ref{tab:BHs}, there is a
significant ``spin up'' relative to the progenitor
angular velocity~\cite{Shibata:1999yx}. The angular frequency increases by
approximately $35$--$49\%$ across all configurations that undergo
collapse. 

These results are shown in Fig.~\ref{fig:delta_om_jm2}, where we plot
the relative increase in the angular velocity against the relative
change in the dimensionless angular momentum. We find that all
configurations undergo a reduction in angular momentum. This behavior
is primarily due to the angular momentum carried away by the
gravitational waves emitted during the collapse, as well as by the
matter that is expelled from the system and remains outside the black
hole. The effect is particularly pronounced for the strongly
magnetized configurations, which consistently exhibit a larger angular
momentum reduction, reaching almost $8\%$ in the case of s2b2. As we
shall discuss next, this enhanced loss is associated with the fact
that strongly magnetized configurations eject significant amounts of
matter that are accelerated to large distances (${\sim} 353.9$~km) and
are therefore not subsequently captured by the black hole, preventing
their angular momentum from contributing to the final black hole spin. 


\subsection{Magnetically-driven baryon winds}
\label{sec:sub:winds}

\begin{figure*}[t]
    \centering
    \includegraphics[width=0.9\textwidth]{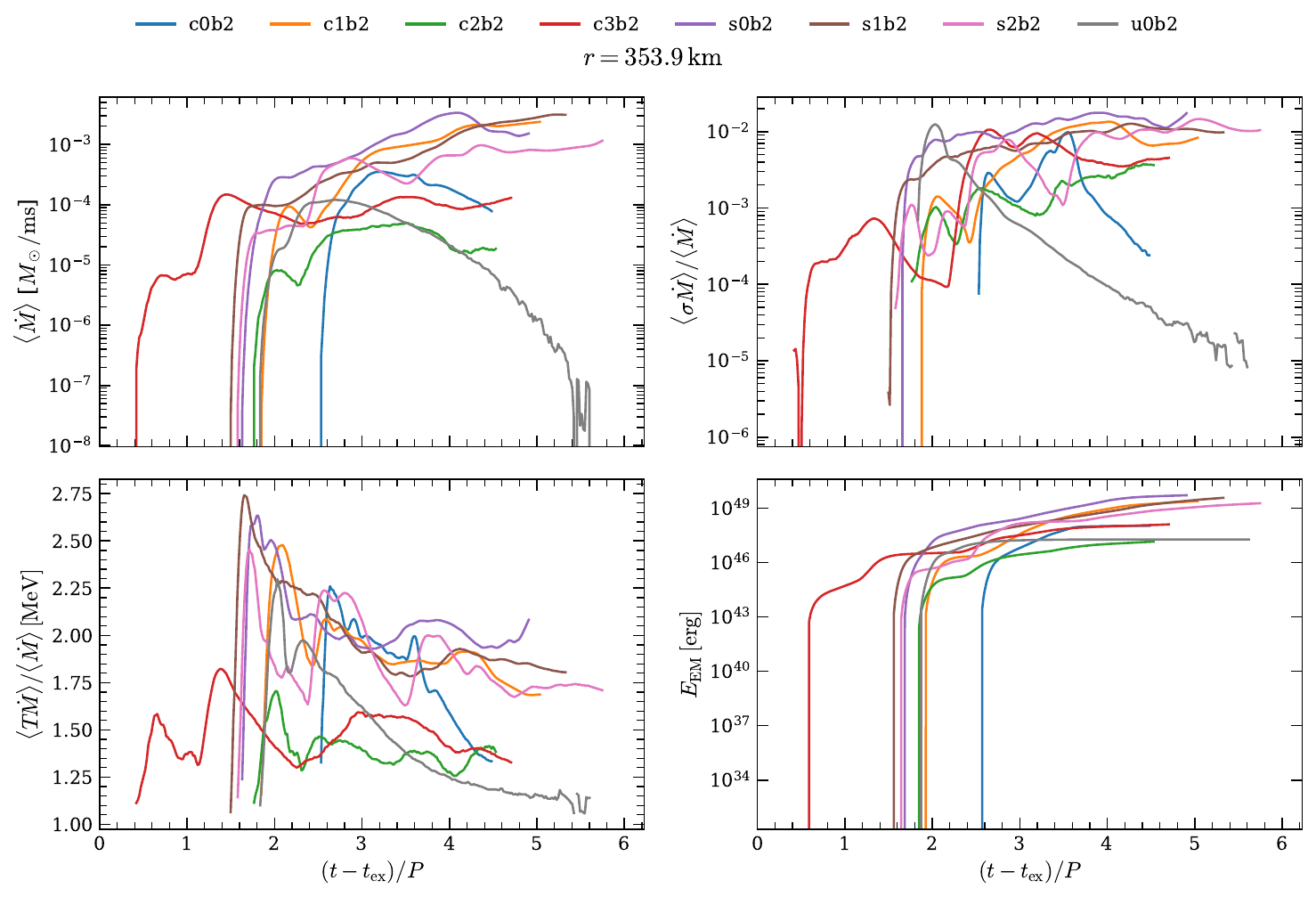}
    \caption{Mass outflow rate, magnetization, temperature and cumulative EM energy measured on ejecta surfaces at an extraction radii of $r=353.9$ km for all simulated configurations. The time-delay is defined as $t_{\rm ex}=r/v_{\infty,\rm max}$, where $v_{\infty,\rm max}$ is the maximum non-zero velocity of the ejecta.}
    \label{fig:mdot_sigma}
\end{figure*}

\begin{figure*}[t]
    \centering
    \includegraphics[width=0.9\textwidth]{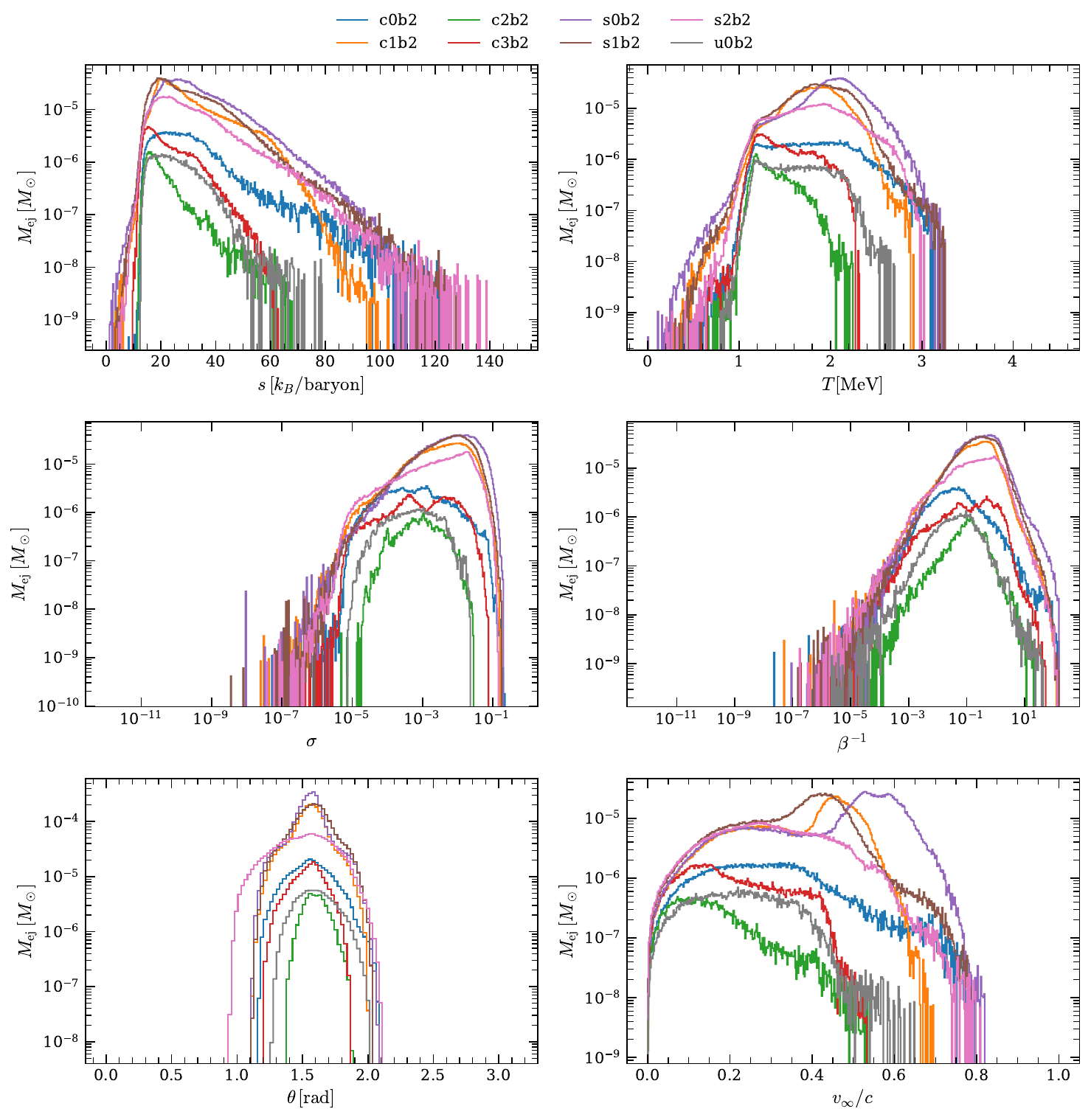}
    \caption{Ejecta histograms of the strongly magnetized configurations. On the y-axis we show the ejected mass and and on the x-axis we show (from left to right, top to bottom) the entropy, temperature, magnetization, ratio of the magnetic and plasma pressures, vertical angle and the asymptotic velocity.}
    \label{fig:histo}
\end{figure*}

The outer layers of the star develop MRI-unstable regions that are
accompanied by magnetic field amplification and the onset of
magnetically driven outflows. This temporal and spatial association
indicates that the growth of magnetic fields in the outer layers,
together with the associated magnetic stresses, contributes to the
development of the observed outflows.

In Figure~\ref{fig:mdot_sigma}, we show the instantaneous baryon mass
flux, magnetization $\sigma=B^2/(W\rho\sqrt{\gamma})$, temperature and
cumulative EM energy measured at an extraction sphere of coordinate
radius $r=353.9$ km for the b2 configurations. Here, the cumulative
electromagnetic energy is defined as the time integral of the Poynting
flux luminosity,
\begin{equation}
E_{\rm EM}(t) = \int_{t_0}^{t} L_{\rm EM}(t')dt',
\end{equation}
where $L_{\rm EM}$ is the instantaneous electromagnetic
(Poynting-flux) luminosity through the extraction sphere. At this
surface, material is flagged as unbound according to the Bernoulli
criterion, $hu_t<-1$, provided that the radial component of the
3-velocity is outward-directed. To facilitate comparison with the
dynamics at the stellar surface, we account for the finite propagation
time of the outflow between the star and the extraction radius. Since
the outflow does not possess a single well-defined propagation
velocity, we estimate the shortest possible travel-time delay using
the maximum non-zero velocity measured among the ejecta,
$t_{\rm ex}=r/v_{\infty,\rm max}$. The quantities shown in the figure
are therefore shifted by this reference time.

The mass flux through the extraction sphere reaches values of a few
$10^{-3}\Mo/\mathrm{ms}$, and the total ejecta mass accumulated over
the simulation is of order ${\sim}10^{-3}\Mo$. The outflows exhibit
typical magnetization values of $\sigma\sim10^{-2}$, indicating that
the ejecta remain matter dominated, although magnetic stresses can
contribute to their acceleration. Temperatures reach a few MeV,
corresponding to ${\sim}10^{10}$--$10^{11}$~K, with substantial
temporal and configuration-dependent variations.

For the instantaneous baryon mass flux, the configurations that collapse the
earliest, namely u0b2, c2b2 and c3b2, exhibit the lowest outflow rates,
reaching ${\lesssim}10^{-4}\Mo\,{\rm ms}^{-1}$. In contrast, the largest
mass fluxes are obtained for s0b2, s1b2 and c1b2, which are either
long lived or undergo collapse at later times. The cumulative
electromagnetic energy shows a similar dependence on the remnant
lifetime, with the long-lived configurations continuing to accumulate
EM energy throughout the simulated evolution, while its growth is
terminated by the collapse of the earliest-collapsing configurations.
The magnetization remains at comparable levels across the
configurations, with no values exceeding $\sigma\sim0.2$. The
temperature exhibits a larger degree of temporal variability, with
peaks of a few MeV and configuration-dependent profiles. In
particular, s1b2 reaches a maximum ejecta temperature of
${\sim}2.4$~MeV at around $(t-t_{\rm ex})/P\sim1.8$, approximately
coincident with the onset of significant differential rotation.
Overall, the strength and duration of the outflows correlate more
clearly with the remnant lifetime than with the instantaneous
magnetization: configurations that remain dynamically active for
longer generally sustain stronger mass outflows and accumulate more
electromagnetic energy.

Figure~\ref{fig:histo} shows mass-weighted ejecta histograms for the b2
configurations. In each subplot, the $y$-axis represents the total
ejected mass accumulated over the full simulation, while the $x$-axis
shows the distribution of the corresponding ejecta properties. For
the entropy per baryon, all distributions peak in the range
$20$--$40\,k_B/$baryon. The temperature distributions peak between $1$
and $2.5$ MeV. The magnetization distributions remain below
$\sigma\sim0.2$, consistent with matter-dominated outflows. The
asymptotic velocity distributions peak at approximately
$v_{\infty}/c\sim0.45$. Longer-lived configurations generically produce later
ejecta with higher entropy, stronger magnetization and larger
asymptotic velocities than the earlier-collapsing configurations.
Ejecta are concentrated around the equatorial plane
($\theta\simeq\pi/2$), indicating they are launched
close to the equator (see Fig.~\ref{fig:rho_xy_xz}). The plasma-$\beta$ distributions show a
broad peak at $\beta^{-1}\sim0.1$--$1$ for most configurations,
indicating that the magnetic and thermal pressures are of comparable
magnitude in a significant fraction of the ejecta. The substantial
spread of the distributions, with $\beta^{-1}$ extending above unity,
also indicates that magnetic pressure can become locally dominant in
part of the outflow. 

A quantitative comparison between the b2 and b1 configurations is not
possible at the extraction radius considered here. In the b1 configurations,
the ejecta are either chocked by the atosmphere environment or the
material remains gravitationally bound, forming a pseudo-disk. 


\section{Conclusion}
\label{sec:con}

We have investigated the dynamical stability and nonlinear evolution
of strongly magnetized SMNS 
close to the mass-shedding limit using three-dimensional
general-relativistic magnetohydrodynamic simulations. We constructed
equilibrium configurations with the finite-temperature SFHo equation
of state and purely poloidal magnetic fields spanning maximum
strengths from ${\sim} 10^{16}$ (b1 series) to ${\sim} 10^{17}\,\mathrm{G}$
(b2 series).

The equilibrium properties of the initial configurations are only
weakly modified by the presence of the magnetic field. However,
dynamical evolution of magnetized configurations can substantially
differ from the non-magnetized case. The initially purely poloidal magnetic
configurations are unstable and rapidly develop a toroidal
component. 
The resulting Maxwell stresses redistribute angular
momentum and drive the development of differential rotation.

In the more strongly magnetized configurations (b2), this process
occurs on Alfv\'en timescales comparable to, or shorter than, the
rotational timescale. These configurations are characterized by Lehnert number
$\Le\gtrsim1$. 
In the more weakly magnetized b1 configurations, for which
$\Le<1$, rotational dynamics remain faster than the magnetic response and
the toroidal field can grow through winding for a longer period before
magnetic instabilities significantly modify its structure.

Magnetic stresses can reduce the rotational support of the central
regions, increasing the central density and triggering
gravitational collapse. We observe this behaviour in b2 configurations
close to the maximum mass threshold and the (unmagnetized) 
marginally stable or unstable branch ($\partial M/\partial\rho_c\lesssim0$).
The remnant black holes have angular frequencies up up 49\% larger
than the initial stars. However, the black hole angular momentum is
smaller compared to the one of the star due to angular momentum
carried away by the gravitational waves and the wind.
The latter is responsible for almost ${\sim}8\%$ of angular momentum
reduction (s2b2).

Our results also indicate that magnetic-field instabilities can dynamically
re-organize the structure of a SMNS, leading to
a new differentially rotating configuration on dynamical timescales. 
Configurations close to the (unmagnetized) stable branch ($\partial M/\partial\rho_c>0$),
do not collapse. Instead, the
stars evolve towards a quasi-stationary state characterized by
substantial differential rotation and a mixed poloidal--toroidal magnetic
field~\footnote{ %
Recent work by \citet{Li:2026ngu}, appeared during the preparation of this manuscript, further suggests that dynamically stable configurations may not be unique.}. 
The resulting differential rotation
then provides an additional source for magnetic winding. The outer
layers of these stars are MRI unstable and produce sustained, and mostly
equatorial, winds with magnetization $\sigma\lesssim 0.2$.


While these outflows are magnetically launched, they remain
matter-dominated at the extraction radii probed here, even though this
radius lies well outside the light-cylinder radius $R_{lc}=
cP/2\pi\approx 25-40$ km for the periods in
Table~\ref{tab:initial_configs}. This is an important distinction from
the canonical, magnetically-dominated $(\sigma \gg 1$) pulsar wind
invoked to explain magnetic dipole spin-down and pulsar-wind-nebula
formation. We caution, however, that our setup is not designed to
resolve a force-free or vacuum magnetosphere, and is therefore not the
ideal tool to test such a magnetically dominate regime. The outflows
identified here are better interpreted as magnetically-driven baryonic
mass ejection, relevant to BNSM remnant enrichment and to
magnetar-flare like ejection events, rather than as direct evidence
for a relativistic, Poynting-flux-dominated pulsar-like wind.
While our evolutions focus on initially cold stars, similar simulations of
``hot'' configurations with neutrino heating effects suggest that
equatorial winds can confine and collimate polar highly-magnetized
ultra-relativistic outflows, with possible implications for
millisecond proto-magnetars and merger remnants~\cite{Desai:2026llu}.

Binary neutron star mergers can produce massive neutron-star remnants,
including long-lived remnants for sufficiently low total binary masses
\cite{Bernuzzi:2020tgt,Radice:2020ddv}. These remnants are expected to
be differentially rotating at formation, strongly magnetized, and
composed of hot, neutrino-trapped matter during the early post-merger
evolution \cite{Perego:2018rht,Espino:2023dei}. They also possess
excess angular momentum and gravitational mass relative to SMNS
configurations with similar baryon mass
\cite{Radice:2018xqa}. Although our models do not capture all of the
physical ingredients relevant to BNSM remnants, they provide a useful
framework for exploring aspects of their long-term dynamical
evolution.

The viscous evolution of BNSM remnants proceeds over timescales of
seconds, during which magnetic stresses, neutrino cooling, and viscous
processes can substantially modify their rotational, magnetic, and
thermal structure. While viscous angular-momentum transport is
commonly expected to drive the remnant toward uniform rotation
\cite{Radice:2017zta,Shibata:2017jyf}, our results suggest that a
significant degree of differential rotation may instead persist. In
particular, the strong magnetic fields expected in these remnants can
reorganize the rotational structure and establish a quasi-stationary,
differentially rotating state on a timescale set by the Lehnert
number. The relevance of this mechanism to realistic BNSM remnants
will, however, depend on whether the magnetic and rotational
timescales remain sufficiently short compared with the timescales
associated with neutrino cooling and viscous angular-momentum
transport. Establishing a more quantitative connection between the
dynamical reorganization observed in our simulations and the secular
evolution of BNSM remnants will require a treatment of magnetic-field
dissipation, neutrino cooling, and long-term angular-momentum
transport that lies beyond the scope of the present simulations.  

A tempting extension of this picture concerns the extended emission
(EE) phase observed in a subset of short gamma-ray bursts, where a
hard initial pulse is followed by softer emission lasting up to
${\approx} 10-100$~s~\cite{Gehrels:2006tk,Dinatolo:2026hef}, which can be attributed to prolonged
activity of a differentially rotating central engine. In our
simulations, the reorganization timescale of the field and rotation
profile is set by the competition between the Alfvén and rotational
timescales, encoded in the Lehnert number: stronger fields and faster
rotation lead to a rapid (over a few periods) transition to a new
quasi-stationary state. Extrapolating to weaker fields at fixed
Lehnert number, the absolute Alfvén and rotational timescales would
increase, potentially extending the reorganization timescale towards
EE-relevant durations. The overall values for the energy emitted by
the system roughly match those expected from the EE phase.
We caution, however, against a quantitative extrapolation of this
kind. The regime explored here is separated from the EE timescale by
two or three orders of magnitude in field strength and rotation
period, and establishing whether the mechanism identified here
persists on the secular timescales relevant to EE will require
simulations extending well beyond the dynamical/Alfv\'en timescale, as
well as the inclusion of viscous and neutrino-cooling physics not
considered in this work.


Finally, as already discussed, several limitations of the present study should be kept in mind. Our
simulations follow the early dynamical evolution over only a few
rotational periods, and the initial data are restricted to purely
poloidal magnetic fields. 
Extending this study to longer timescales, higher
resolutions, and more realistic initial magnetic geometries and a better treatment of a highly magnetized atmosphere will be
necessary to determine the long-term evolution of the quasi-stationary
configurations identified here. 
The evolution of initial data for rotating configuration with mixed
poloidal--toroidal magnetic fields appears particularly important for
determining whether the configuration found in this work may represent
stable configurations.


\begin{acknowledgments}
  We thank Nicolo Bucciantini, Aurora Capobianco, Luca Del Zanna and Raj Kishor Joshi for help and discussions during the development of this project. 

  JLRM is supported by the research grant PID2020-118635GB-I00 from
  the Spain-Ministerio de Ciencia e Innovaci\'{o}n and by the Grant
  ED431F 2025/04 of the Galician CONSELLERIA DE EDUCACION, CIENCIA,
  UNIVERSIDADES E FORMACION PROFESIONAL. JLRM also acknowledges
  support from the Xunta de Galicia through file number ED481A-2026-253
  (Programa de axudas á etapa predoutoral). IGFAE is supported by the 
  Mar\'ia de Maeztu Unit of Excellence grant CEX2023-001318-M, funded by MICIU/AEI/10.130939/501100011033.
  SB and BH acknowledge support for the MERLIN project. MERLIN is
  funded by the Deutsche Forschungsgemeinschaft (DFG) and the Narodowe
  Centrum Nauki (NCN) OPUS-LAP grant number 2022/47/I/ST9/01494, under
  the EU weave initiative.
  SB acknowledges support by the DFG project ``Magnetfelddynamik in
  Neutronensternen Sternen'' MERLIN (MERLIN; BE 6301/6-1
  Projektnummer: 524726453).
  SB acknowledges support by the EU Horizon under ERC Consolidator Grant, no. InspiReM-101043372.
  SB and MJ acknowledge support from the Deutsche Forschungsgemeinschaft, DFG, project MEMI number BE 6301/2-1.
  WC is supported by a Royal Society University Research Fellowship  (URF\textbackslash R1\textbackslash 261458).


  The authors acknowledge the Gauss Centre for Supercomputing
  e.V. (\url{www.gauss-centre.eu}) for funding this project by providing
  computing time on the GCS Supercomputer SuperMUC-NG at LRZ
  (allocations {\tt pn76li} and {\tt pn67xo}. The authors
  acknowledge HLRS for funding this project by providing access to the
  supercomputer HPE Apollo Hawk under the grant number INTRHYGUE/44215
  and MAGNETIST/44288.

  Computations were also performed on the ARA cluster at Friedrich
  Schiller University Jena and on the {\tt Tullio} INFN cluster at INFN
  Turin. The ARA cluster is funded in part by DFG grants INST 275/334-1
  FUGG and INST 275/363-1 FUGG, and ERC Starting Grant, grant agreement
  no. BinGraSp-714626. 
\end{acknowledgments}


\appendix

\section{Convergence tests}
\label{app:convergence}

We have verified the numerical convergence of our results by comparing several representative configurations at different resolutions. As described in Sec.~\ref{sec:method}, all results presented in this work are obtained using a base resolution of 128 grid points with seven refinement levels. We find that a lower base resolution of 64 grid points with the same seven refinement levels is already sufficient to capture the main dynamical features of the evolution. The good agreement between the two resolutions provides additional evidence for the robustness of our results and supports the adequacy of the adopted numerical setup.

\begin{figure*}[t]
    \centering
    \includegraphics[width=.9\textwidth]{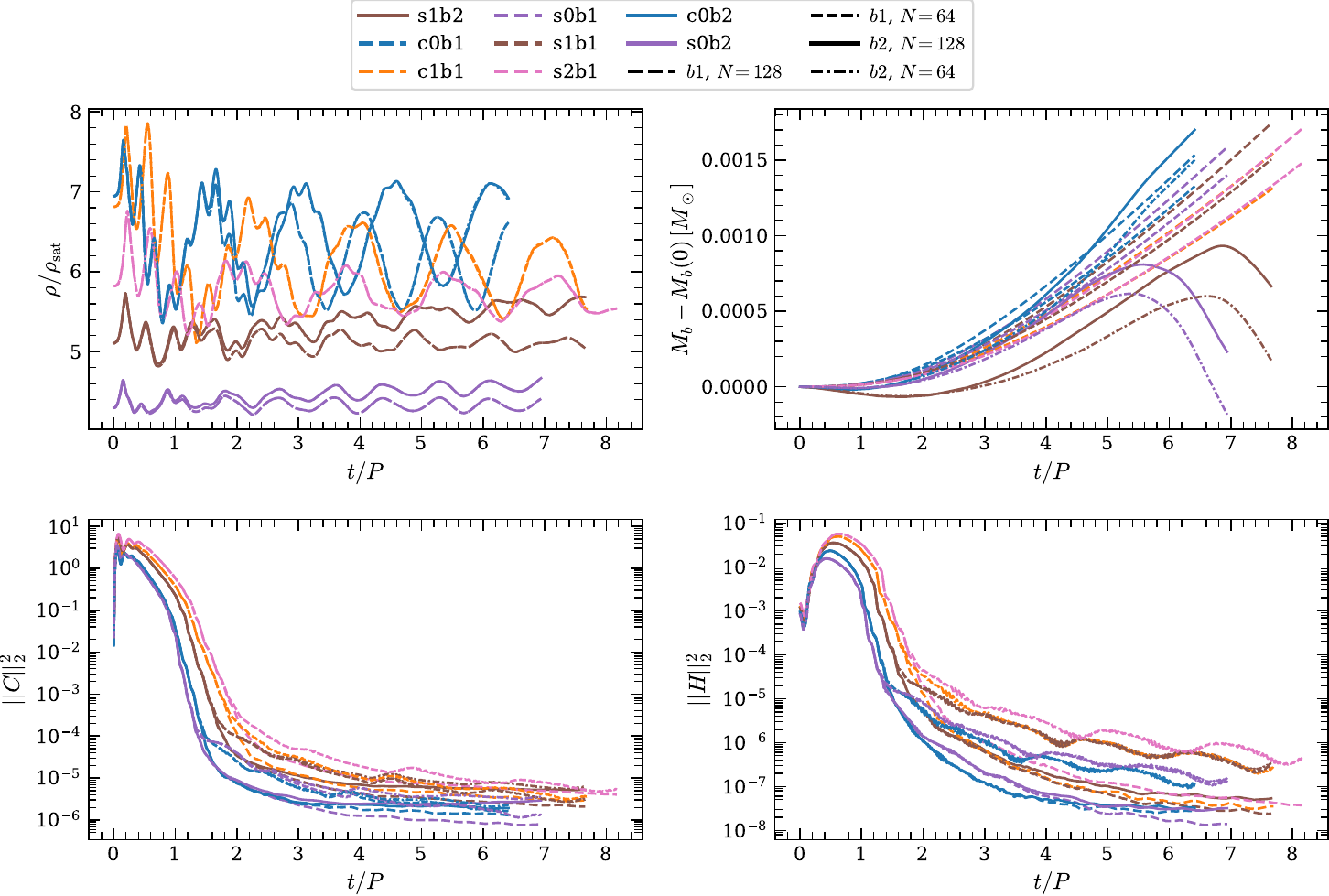}
    \caption{Convergence tests comparing three representative simulations with their corresponding higher-resolution counterparts. Shown are the time evolution of the density, baryonic mass, and curvature-related quantities ($||C||^2_{2}$ and $||H||^2_{2}$).}
    \label{fig:rho_check}
\end{figure*}

\begin{figure*}[t]
    \centering
    \includegraphics[width=.9\textwidth]{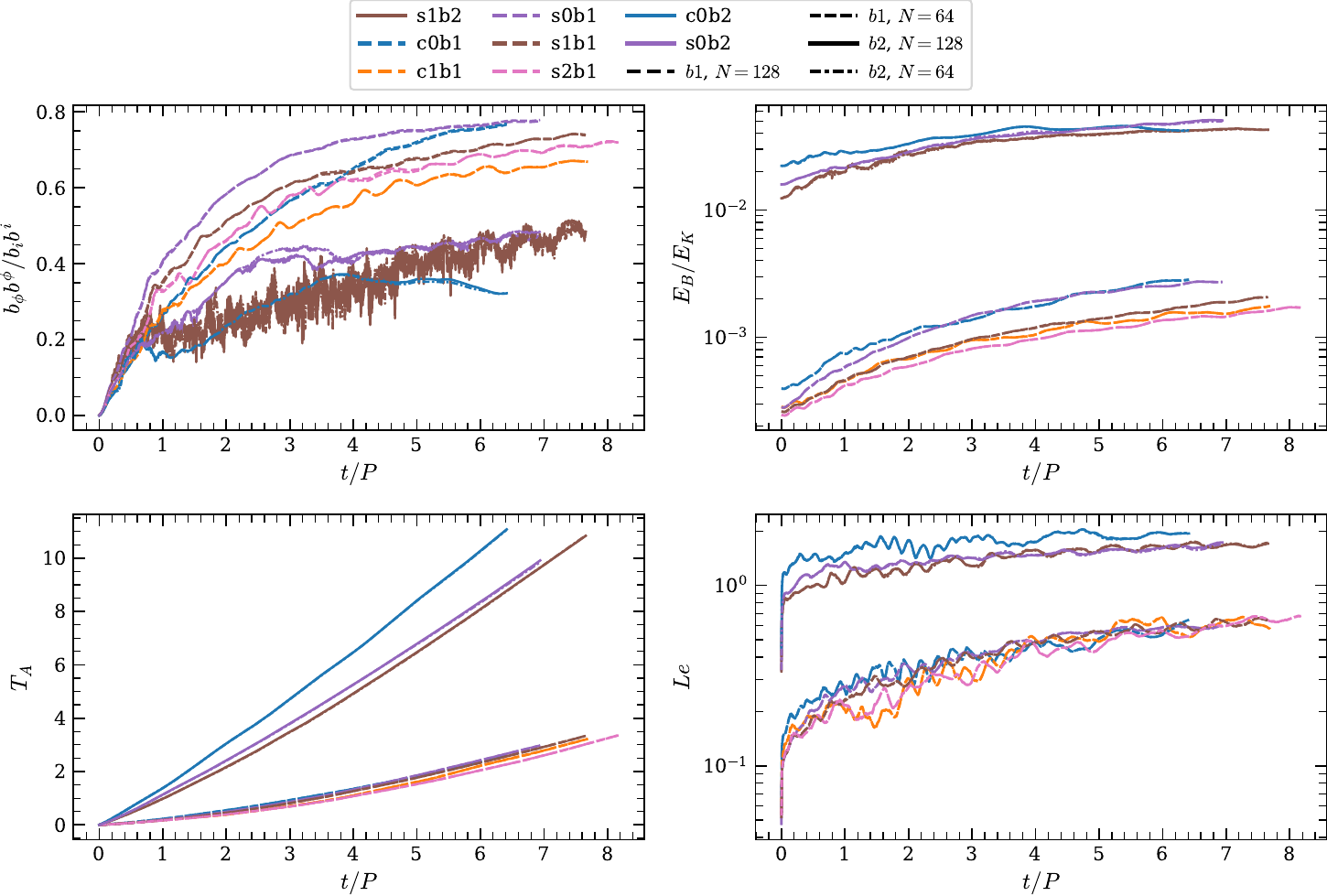}
    \caption{Convergence tests comparing three representative magnetized simulations with their corresponding higher-resolution counterparts. Shown are the maximum toroidal magnetic field ratio, the magnetic-to-kinetic energy ratio, the Alfv\'en crossing time, and Lehnert number.}
    \label{fig:mag_check}
\end{figure*}

\begin{table}[t]
\centering
\caption{Relative change between low and high resolutions for the BH properties in Tab.~\ref{tab:BHs}. Here, $\Delta X = 100(X_{64}-X_{128})/X_{128}$, where $X_{128}$ denotes the higher-resolution result and $X_{64}$ the lower-resolution result. The columns ``B.N.'' and ``M.L.'' indicate the base name and the magnetic label of each configuration.}
\begin{tabular}{|c|c|c|c|c|c|c|}
\hline
B.N. & M.L. & $\Delta(t_{\mathrm{col}}/P)$[\%] & $\Delta M_{\mathrm{BH}}$[\%] & $\Delta(S_H/M_{\mathrm{BH}}^2)$[\%] & $\Delta\Omega_H$[\%] \\
\hline
\multirow{1}{*}{c1} & b2 & +3.368 & -0.177 & +0.305 & +0.544 \\
\hline
\multirow{3}{*}{c2} & b0 & +3.182 & +0.021 & +0.103 & +0.136 \\
\cline{2-6}
 & b1 & +4.382 & +0.012 & +0.129 & +0.153 \\
\cline{2-6}
 & b2 & +0.524 & -0.803 & +1.604 & +2.873 \\
\hline
\multirow{3}{*}{c3} & b0 & +0.515 & -0.024 & -0.054 & -0.048 \\
\cline{2-6}
 & b1 & +0.529 & -0.018 & -0.096 & -0.111 \\
\cline{2-6}
 & b2 & +0.602 & +0.305 & -0.541 & -1.027 \\
\hline
\multirow{1}{*}{s2} & b2 & +0.623 & -1.806 & +3.662 & +6.801 \\
\hline
\multirow{3}{*}{u0} & b0 & +0.000 & +0.010 & +0.076 & +0.095 \\
\cline{2-6}
 & b1 & +0.000 & +0.010 & +0.077 & +0.096 \\
\cline{2-6}
 & b2 & +0.000 & -1.978 & +4.230 & +7.954 \\
\hline
\end{tabular}
\label{tab:BHs_resol}
\end{table}

As shown in Figs.~\ref{fig:rho_check} and~\ref{fig:mag_check}, differences between simulations and their lower-resolution counterparts are generally negligible during the bulk evolution. Noticeable deviations only appear after black hole formation, where gauge effects and the loss of a well-defined fluid interior make many history variables less informative.

We have also compared global properties of the resulting black holes and find good overall agreement between resolutions. As shown in Tab.~\ref{tab:BHs_resol}, the final black hole mass generally exhibits relative differences below $2\%$, while the dimensionless spin and horizon angular frequency show differences of up to $4.2\%$ and $8.0\%$, respectively. The largest discrepancies in these spin-related quantities occur for the configurations with the strongest magnetic field ($b2$), particularly for the u0b2 configuration. This suggests that the spin-related properties are more sensitive to resolution in these strongly magnetized cases. Nevertheless, the overall agreement remains good, especially considering the different dynamical timescales and magnetic-field strengths involved. Differences between resolutions are generally small during the bulk evolution. The collapse time exhibits somewhat larger variations, reaching approximately $4.4\%$ in the c2b1 configuration. This sensitivity is expected to be more pronounced near the stability threshold, where small perturbations can affect the onset of collapse. We emphasize that the collapse time is not a strictly well-posed observable in dynamical spacetimes and should therefore be regarded as an indicative quantity.

Finally, we have assessed the impact of the ``artificial'' atmosphere by reducing the floor density to $\rho_{\rm atm}=1860~\mathrm{g/cm^3}$ to test whether the relatively high atmosphere required by the strongly magnetized runs affects the dynamics. We find that its influence is smaller than the differences induced by the resolution changes. In particular, the collapse time changes by only ${\sim} 0.5\%$, while the conserved baryonic mass changes only at approximately the third to fourth significant digit, corresponding to deviations of at most ${\sim} 0.1\%$. These results indicate that the chosen atmosphere prescription does not introduce significant or systematic effects on the neutron-star evolution. Nevertheless, an improved treatment of the atmosphere would be desirable, particularly for strongly magnetized configurations, as it would allow for lower atmosphere densities while maintaining numerical stability and could therefore further improve mass conservation. In addition, the $||H||_2^2$- and $||C||_2^2$-norms shown in Fig.~\ref{fig:rho_check} decrease throughout the evolution, providing a further indication that the numerical evolution remains well behaved.


\bibliography{references.bib}

\end{document}